\documentclass[lettersize,journal]{IEEEtran}
\usepackage{amsmath,amsfonts}
\usepackage{algorithmic}
\usepackage{algorithm}
\usepackage{array}
\usepackage[caption=false,font=normalsize,labelfont=sf,textfont=sf]{subfig}
\usepackage{textcomp}
\usepackage{stfloats}
\usepackage{url}
\usepackage{verbatim}
\usepackage{graphicx}
\usepackage{cite}
\usepackage{multirow}
\usepackage{makecell}
\usepackage{threeparttable}
\usepackage[normalem]{ulem}
\begin{document}

\title{Vesper: A Compact and Effective Pretrained Model for Speech Emotion Recognition}
\author{Weidong Chen,~\IEEEmembership{Student Member,~IEEE,} Xiaofen Xing,~\IEEEmembership{Member,~IEEE,} Peihao Chen \\ and Xiangmin Xu,~\IEEEmembership{Senior Member,~IEEE}
\thanks{Weidong Chen and Xiaofen Xing are with the School of Electronic and Information Engineering, South China University of Technology, Guangzhou 510640, China (e-mail: eewdchen@mail.scut.edu.cn; xfxing@scut.edu.cn).}
\thanks{Peihao Chen is with the School of Software Engineering, South China University of Technology, Guangzhou 510640, China (e-mail: phchencs@gmail.com).}
\thanks{Xiangmin Xu is with the School of Future Technology, South China University of Technology, Guangzhou 511442, China, and also with Pazhou Laboratory, Guangzhou 510330, China (e-mail: xmxu@scut.edu.cn).}
\thanks{Corresponding authors: Xiangmin Xu and Xiaofen Xing.}
}


\maketitle

\begin{abstract}
This paper presents a paradigm that adapts general large-scale pretrained models (PTMs) to speech emotion recognition task. Although PTMs shed new light on artificial general intelligence, they are constructed with general tasks in mind, and thus, their efficacy for specific tasks can be further improved. Additionally, employing PTMs in practical applications can be challenging due to their considerable size. Above limitations spawn another research direction, namely, optimizing large-scale PTMs for specific tasks to generate task-specific PTMs that are both compact and effective. In this paper, we focus on the speech emotion recognition task and propose an impro\uline{V}ed \uline{e}motion-\uline{s}pecific \uline{p}retrained encod\uline{er} called Vesper. Vesper is pretrained on a speech dataset based on WavLM and takes into account emotional characteristics. To enhance sensitivity to emotional information, Vesper employs an emotion-guided masking strategy to identify the regions that need masking. Subsequently, Vesper employs hierarchical and cross-layer self-supervision to improve its ability to capture acoustic and semantic representations, both of which are crucial for emotion recognition. Experimental results on the IEMOCAP, MELD, and CREMA-D datasets demonstrate that Vesper with 4 layers outperforms WavLM Base with 12 layers, and the performance of Vesper with 12 layers surpasses that of WavLM Large with 24 layers.

\end{abstract}


\section{Introduction}

\IEEEPARstart{P}{revalent} in the realm of artificial intelligence are large-scale pretrained models (PTMs) comprising hundreds of millions and, in some cases, billions of parameters, which have exhibited remarkable performance across various tasks \cite{survey_speech, survey}. PTMs are recognized as key components of artificial general intelligence due to their ability to solve multiple tasks simultaneously \cite{gpt-3}. PTMs are pretrained using extensive amounts of unlabeled data and then generalized to specific tasks to achieve human-like performance. Among them, PTMs such as BERT \cite{bert}, GPT-3 \cite{gpt-3}, and CLIP \cite{clip} have achieved exceptional results across a wide range of natural language processing and computer vision tasks. Similarly, in the speech signal processing domain, PTMs such as wav2vec \cite{wav2vec}, wav2vec 2.0 \cite{wav2vec2}, HuBERT \cite{hubert}, and WavLM \cite{wavlm} have pushed the boundaries of various speech tasks and delivered promising performance.

\begin{figure}[t]
\centering
\includegraphics[width=0.95\linewidth]{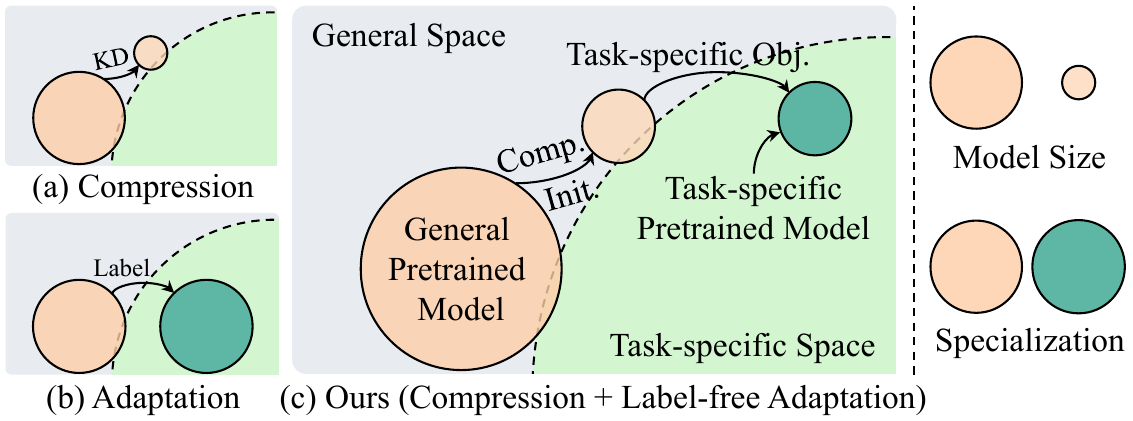}
\caption{(a) Compressing large-scale pretrained model by knowledge distillation. (b) Adapting a large-scale pretrained model to a downstream task by labeled task-related data. (c) Our pipeline simultaneously applies compression and label-free adaptation to generate a task-specific pretrained model. KD, Comp., Init., and Obj. stand for knowledge distillation, compression, initialization, and objective, respectively. 
Circles of different colors represent models specialized for different spaces or tasks.
}
\label{intro}
\end{figure}

The triumph of large-scale PTMs has sparked researchers' interest in acquiring greater amounts of unlabeled data to pretrain larger models with billions of parameters \cite{gpt-3, llama, Palm}. However, current PTMs are pretrained in a task-agnostic manner because they are designed to capture general representations that can be applied to various tasks. Consequently, there is always room for improvement on specific tasks. Moreover, the substantial computational and storage resources required for large-scale PTMs make them challenging to apply in practical scenarios.

The future development of PTMs will not only focus on creating large-scale general PTMs but also explore another research direction, which is generating task-specific PTMs by additional pretraining of general PTMs using task-specific objectives. Task-specific PTMs are expected to exhibit the following features: (1) Task-specific pretraining enables task-specific PTMs to capture critical task-specific representations from the general representations learned from massive unlabeled data. (2) Task-specific PTMs are lightweight because they focus on one specific task. In summary, task-specific PTMs should be compact and effective.

Current studies \cite{ser_with_pre, qa_with_pre, adapter, as_with_pre, distilling, adversarial, DistilBERT, AdaBERT} on using general PTMs for specific tasks have explored two main directions: (1) As shown in Fig.~\ref{intro}(a), researchers have attempted to reduce the model size of PTMs to overcome the latency and capacity constraints. In particular, the knowledge distillation technique is applied to transfer knowledge from a large-scale PTM to a compact model \cite{distilling, DistilBERT, AdaBERT}. Despite model compression, knowledge distillation fails to incorporate the characteristics of specific tasks, and thus, the compact model is still task-agnostic, leading to an incompatibility between the general pretrained objectives and specific downstream tasks. (2) As shown in Fig.~\ref{intro}(b), the large-scale PTM is directly adapted to downstream tasks by fine-tuning on labeled task-related corpora \cite{qa_with_pre, adapter}. However, acquiring sufficient labeled data for downstream tasks is difficult and costly. Additionally, the model size of the PTM remains unchanged, which limits practical deployment. To overcome these limitations, we integrate compression and label-free adaptation into a single pipeline to generate a task-specific PTM that is both compact and effective, as illustrated in Fig.~\ref{intro}(c).

We focus on the speech emotion recognition task and present an impro\uline{V}ed \uline{e}motion-\uline{s}pecific \uline{p}retrained encod\uline{er} called Vesper. Vesper is pretrained further on the basis of WavLM \cite{wavlm} in a self-supervised manner with an emotion-related training strategy. It is initialized with the parameters of WavLM and achieves compression by reducing the number of employed layers. Masked prediction is applied as the label-free training objective. To improve Vesper's sensitivity to emotional information, we design a novel emotion-guided masking strategy. In particular, we utilize the energy of the speech signal to identify regions that contain emotional information with high probability and apply masking only within these regions. Recent research \cite{12layer} has shown that the shallow layers of speech PTMs tend to capture acoustic features, while the deep layers tend to capture semantic features. As both acoustic and semantic features are crucial for emotion recognition \cite{A_and_S, A_and_S_2}, we employ hierarchical self-supervision to separately supervise the shallow and deep layers of Vesper. To enrich acoustic information in the deep layer output, we propose cross-layer self-supervision to make the output representation more informative and balanced.

The contributions of this paper can be summarized as follows:
\begin{itemize}
\item We propose a new pipeline that generalizes large-scale pretrained models on speech emotion recognition task by compression and emotion-specific adaptation. We hope that the pipeline inspires researchers to generate compact and effective pretrained models for various speech tasks.
\item We focus on the speech emotion recognition task and propose an emotion-specific pretrained encoder called Vesper. To enhance Vesper's sensitivity to emotional information, we introduce an emotion-guided masking strategy during pretraining, leveraging the energy of the input speech signal to identify the regions that need masking. We also propose a hierarchical self-supervision approach to enhance Vesper's capability to capture both acoustic and semantic information and present a cross-layer self-supervision approach to improve the informativeness and balance of the final output representation.
\item We evaluate Vesper on three widely used emotion recognition datasets, namely, IEMOCAP, MELD, and CREMA-D. Experimental results demonstrate that our Vesper with 4 layers outperforms WavLM Base with 12 layers and that Vesper with 12 layers outperforms WavLM Large with 24 layers. Our code and the pretrained Vesper are publicly available at \url{https://github.com/HappyColor/Vesper}.
\end{itemize}

The remainder of this paper is organized as follows: In Section \uppercase\expandafter{\romannumeral2}, we provide a literature review on large-scale pretrained models and their applications. In Section \uppercase\expandafter{\romannumeral3}, we elaborate on the proposed Vesper. In Section \uppercase\expandafter{\romannumeral4}, we describe the experimental corpora and setup in detail. In Section \uppercase\expandafter{\romannumeral5}, we present our experimental results and analyses. Finally, we present our conclusions in Section \uppercase\expandafter{\romannumeral6}.

\section{Related Work}

In this section, we present an overview of large-scale pretrained models in artificial intelligence and review the various ways researchers employ them in downstream tasks. Subsequently, we systematically introduce Transformer and WavLM, as these two frameworks are the basis of Vesper.

\subsection{Large-Scale Pretrained Models in Artificial Intelligence}

In recent years, the development of large-scale pretrained models has revolutionized the field of artificial intelligence. PTMs leverage vast amounts of unlabeled data and computational power to learn general representations for various tasks. BERT \cite{bert}, RoBERTa \cite{roberta}, T5 \cite{t5}, and GPT-3 \cite{gpt-3} are the most popular PTMs in the field of natural language processing. They have achieved remarkable results in text classification tasks such as named entity recognition \cite{ner_with_pre} and question answering \cite{qa_with_pre}, as well as generative tasks such as machine translation \cite{mt_with_pre} and abstractive summarization \cite{as_with_pre}. GPT-3 \cite{gpt-3} sheds new light on artificial general intelligence and has spawned numerous practical applications. PTMs have also made significant strides in the field of computer vision, ushering in a new era. Vision PTMs such as MoCo \cite{moco}, ViT \cite{vit}, simCLR \cite{simclr}, and video models \cite{visual_1, visual_2} capture high-level visual features to generalize on diverse visual tasks, including object detection \cite{od_with_pre}, image segmentation \cite{is_with_pre}, and image captioning \cite{ic_with_pre}. ViT \cite{vit} has emerged as a highly influential and widely adopted architecture in computer vision. Vision-language pretrained models jointly learn universal vision and language features by pretraining on large-scale image-text pairs. SAM \cite{sam}, CLIP \cite{clip}, Flamingo \cite{flamingo}, and DALL-E \cite{dall-e} achieve strong performance in cross-modal matching, cross-modal reasoning, and cross-modal generation. Various PTMs in the speech domain have also been proposed, including wav2vec~2.0 \cite{wav2vec2}, HuBERT \cite{hubert}, and WavLM \cite{wavlm}. Equipped with the above speech PTMs, researchers have showcased their superior performance in a wide range of speech-related tasks, including automatic speech recognition \cite{wav2vec2, wavlm, asr_pre}, speech enhancement \cite{se_with_pre, se_with_pre_2}, and speech emotion recognition \cite{ser_with_pre, dst, shiftser, tac_1}.

\subsection{Utilizing Pretrained Models for Specific Tasks}

After large-scale PTMs became available, many researchers started exploring their potential applications for various specific tasks. They have mainly focused on the following three directions.

\subsubsection{Transfer Learning}

One key advantage of large-scale PTMs is their remarkable capacity for transfer learning. By leveraging the universal knowledge learned from large-scale unlabeled data, PTMs can be used as a powerful starting point for a new task with limited labeled data \cite{qa_with_pre, as_with_pre, od_with_pre, finetune}. Fine-tuning is one of the most widely used implementations of transfer learning. Cao \textit{et al.} \cite{qa_with_pre} initialized a model with the pretraining weights of BERT \cite{bert} and subsequently performed fine-tuning on the target corpus. Fabbri \textit{et al.} \cite{as_with_pre} created dataset-specific unlabeled data for fine-tuning by leveraging the characteristics of the target dataset to improve zero-shot learning of the model. The transferability of pretrained representations reduces the need for extensive labeled data in specific tasks. However, updating the PTM is resource-intensive due to the substantial model size of the PTM. The PTM is also difficult to deploy in resource-constrained real-world applications.

\subsubsection{Knowledge Distillation}

PTMs can serve as ``teachers" to transfer their knowledge to a smaller ``student" model using the knowledge distillation technique \cite{DistilBERT, fastbert, distilling, adversarial, DAAE, LanSER}. Generally, the objective function employed for this purpose is the Kullback-Leibler (KL) divergence loss, which constrains the probability distribution of the student model to mimic the teacher's prediction.
Typically, Sanh \textit{et al.} \cite{DistilBERT} introduced a distilled version of BERT known as DistilBERT, which achieved 97\% of the performance of BERT while utilizing only 60\% of the model size. Liu \textit{et al.} \cite{fastbert} adopted a self-distillation mechanism and achieved promising results across twelve datasets.
Although the model has been successfully compressed, the compact student model is still a general model that lacks specificity for specific tasks. To obtain a task-specific PTM, labeled data from the target domain are necessary. Zhang \textit{et al.} \cite{adversarial} utilized adversarial samples to augment limited labeled data from the target domain to enhance task-specific knowledge transfer.

\subsubsection{Feature Extraction}

PTMs can also serve as feature extractors, where the pretrained model is frozen and utilized to extract acoustic features from raw input data \cite{ksT, rahman, dst, extractor, CA-MSER, Phukan, wu}. The extracted features are fed into a separate classifier or downstream model designed for the given task. Rahman \textit{et al.} \cite{rahman} employed three pretrained CNN models to extract features from chest X-ray images. Chen \textit{et al.} \cite{dst} adopted WavLM \cite{wavlm} to extract the acoustic features from raw audio samples and achieved remarkable results for speech emotion recognition. Even though the downstream model is usually quite small, using PTM as a feature extractor greatly increases the overall size of the system. Furthermore, general representations may not be effectively mapped to a specific space by using the simple classifier only. The task-specific downstream model requires manual design and considerable manual tuning.

\subsection{Revisiting Transformer and WavLM}
Transformer \cite{Transformer} originally consisted of an encoder and decoder. In this paper, we describe the encoder part only since it is what is needed to implement our proposed architecture. The Transformer encoder contains a multihead self-attention (MSA) module and a fully connected feed-forward network (FFN).
The input of Transformer, $x \in \mathbb{R}^{T \times d}$, is an arbitrary sequence, where $T$ indicates the sequence length and $d$ indicates the feature dimension.
The feed-forward network (FFN) consists of two linear projections with a ReLU activation \cite{relu} in between.
The computational process of Transformer encoder is depicted as follows:
\begin{equation}
Tr(x) = FFN(MSA(x))
\label{quation_transformer}
\end{equation}
where $Tr$ represents the Transformer encoder. More details can be found in \cite{Transformer}.

Given that WavLM \cite{wavlm} exhibits state-of-the-art results on the SUPERB benchmark \cite{superb}, we select WavLM as the foundational model in this paper. WavLM \cite{wavlm} is a self-supervised pretrained model built on the Transformer encoder, which is pretrained on several large corpora and learns to encode audio for general purposes. Compared to other pretrained models in the speech domain that focus mainly on phoneme classification and automatic speech recognition tasks, WavLM utilizes a speech denoising objective in addition to masked speech prediction in the pretraining procedure to jointly extract speech content and acoustic information. Benefitting from denoising modeling, WavLM is extended to full stack speech processing tasks. Additionally, WavLM applies gated relative position bias to the Transformer structure to better model the sequential information of the input speech signal. WavLM comes in two versions, Base and Large, with different model sizes and computational complexities. WavLM Base consists of 12 Transformer layers, and WavLM Large consists of 24 Transformer layers. Detailed configurations can be found in \cite{wavlm}. Finally, WavLM is evaluated on the SUPERB \cite{superb} benchmark. The experimental results demonstrate that WavLM achieves promising performance across 14 downstream tasks, including automatic speech recognition, speaker verification, speech separation, and speech emotion recognition.

\section{Proposed Paradigm and Vesper}

\begin{figure}[t]
\centering
\includegraphics[width=0.9\linewidth]{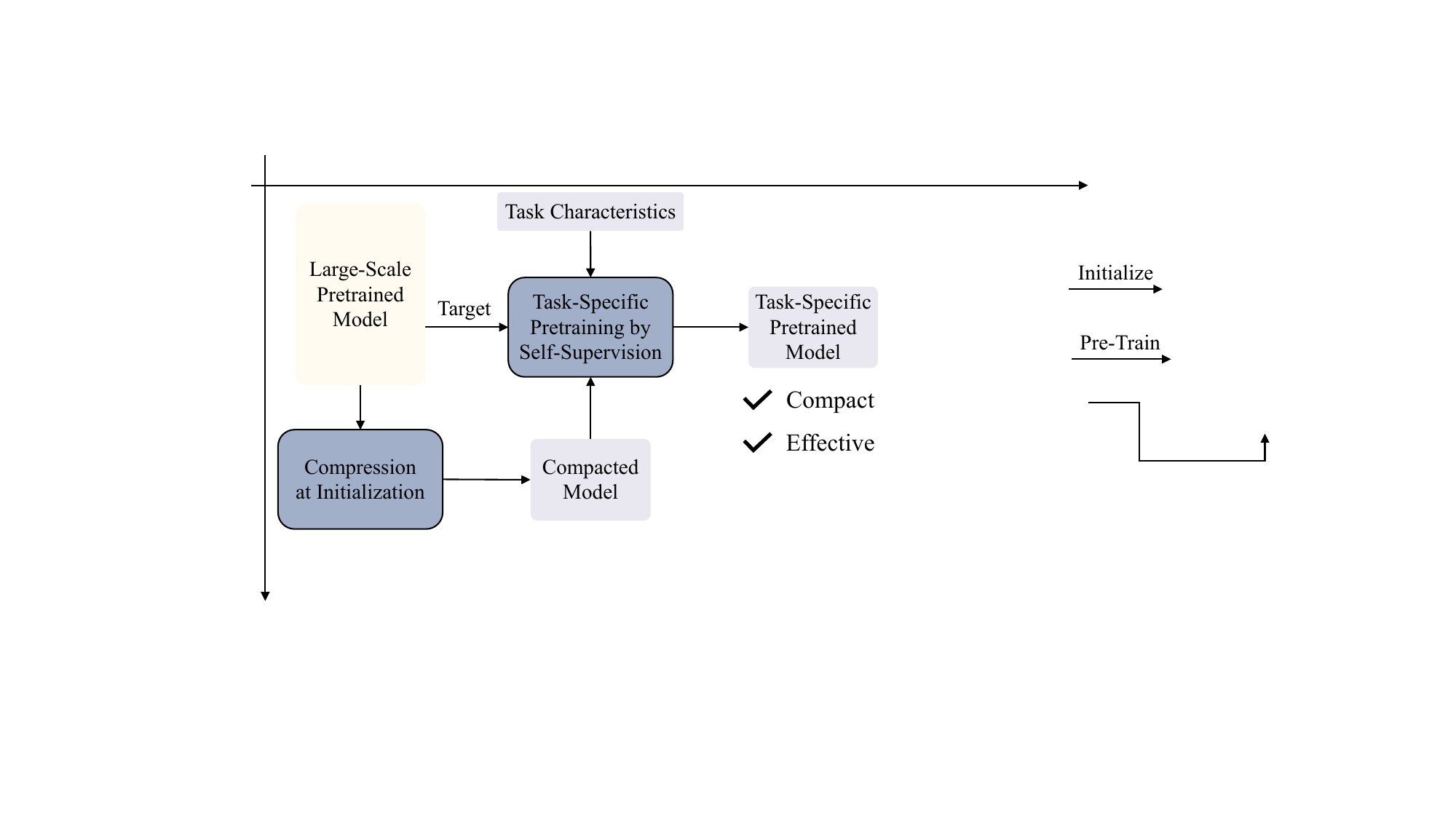}
\caption{The proposed paradigm for generating a task-specific pretrained model that is both compact and effective based on a large-scale pretrained model. The paradigm consists of two steps: compression and task-specific pretraining.}
\label{paradigm}
\end{figure}

As shown in Fig~\ref{paradigm}, the proposed paradigm consists of two main steps: compression at initialization and task-specific pretraining by self-supervision. The first step combines initialization with compression and generates a compact model based on the large-scale pretrained model. In the second step, the compact model is further pretrained by incorporating the characteristics of the downstream task in a self-supervised manner. Eventually, a task-specific pretrained model that is both compact and effective is produced. In this paper, we focus on speech emotion recognition task and create an emotion-specific Vesper following the above paradigm. Vesper is built upon the pretrained WavLM Large model. The model architectures of Vesper and WavLM are identical except for the difference in the number of Transformer layers employed. In the following subsections, to each step involved in building Vesper is comprehensively described.

\subsection{Compression at Initialization}
\label{sec_init}

\begin{figure}[t]
\centering
\includegraphics[width=0.96\linewidth]{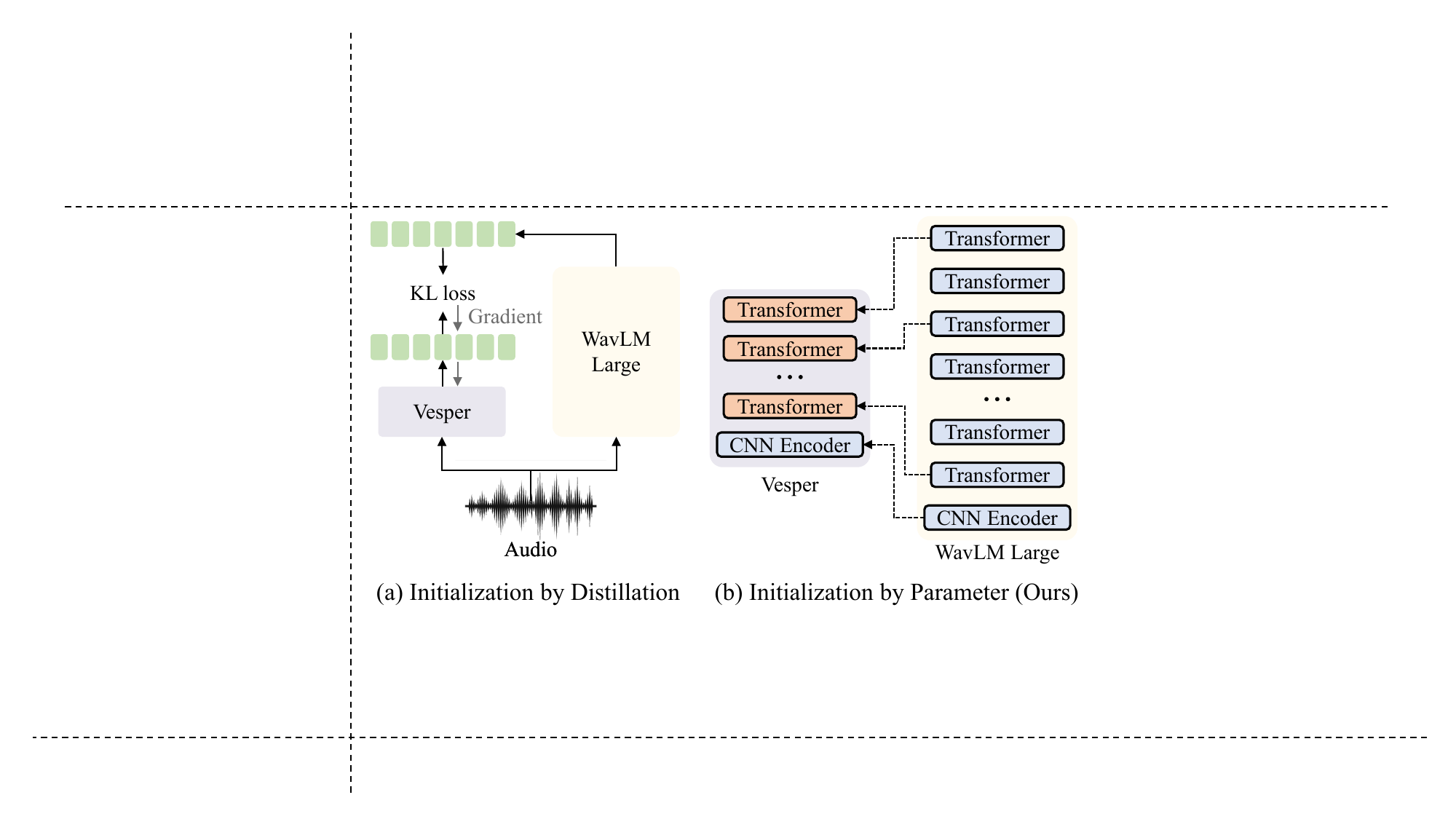}
\caption{Two types of compression approaches. The dashed line in (b) represents copying parameters from WavLM Large directly to Vesper. In this paper, we use approach (b) for initialization.}
\label{init}
\end{figure}

To take full advantage of the knowledge obtained from massive unlabeled speech data in the large-scale pretrained model, current studies apply mainly knowledge distillation techniques to implement knowledge transfer. As illustrated in Fig.~\ref{init}(a), the pretrained WavLM Large model serves as the ``teacher'', while Vesper acts as the ``student''. To train the student model, KL loss is utilized to ensure that the student output mimics the distribution of the teacher output. The distillation operation generates a compact student model that retains general representation and can be further pretrained according to task-specific objectives.

Given that Vesper and WavLM possess the same model architecture, it is worth investigating the possibility of directly initializing Vesper with WavLM's parameters. As shown in Fig.~\ref{init}(b), the CNN encoder in Vesper is directly taken from WavLM. Suppose the numbers of Transformer layers employed in Vesper and WavLM Large are $N$ and $M$, respectively. In this paper, $N$ is much smaller than $M$ for the purpose of compression. We attempt to uniformly extract the Transformer layers from WavLM Large to initialize the Transformer layers in Vesper. In particular, the $i$-th Transformer layer in Vesper is initialized by the parameter of the $(1+\lfloor\frac{M}{N}\rfloor \times (i-1))$-th Transformer layer in WavLM Large, where $i\in[1, N]$; $\lfloor \cdot \rfloor$ rounds numbers down to the nearest integer. In addition to uniform extraction, we try to initialize the Transformer layers in Vesper by uniformly averaging the parameters across the Transformer layers in WavLM Large, which combines the representational capabilities of each layer like ensemble  learning and model fusion.
For example, the parameters of the $(1+\lfloor\frac{M}{N}\rfloor\times(i-1))$-th to the $(\lfloor\frac{M}{N}\rfloor \times i)$-th Transformer layers in WavLM Large are averaged, and the averaged results are used to initialize the $i$-th Transformer layer in Vesper. In such cases, compression and knowledge transfer are implemented simultaneously during model initialization.

\begin{figure*}[th]
\centering
\includegraphics[width=0.9\linewidth]{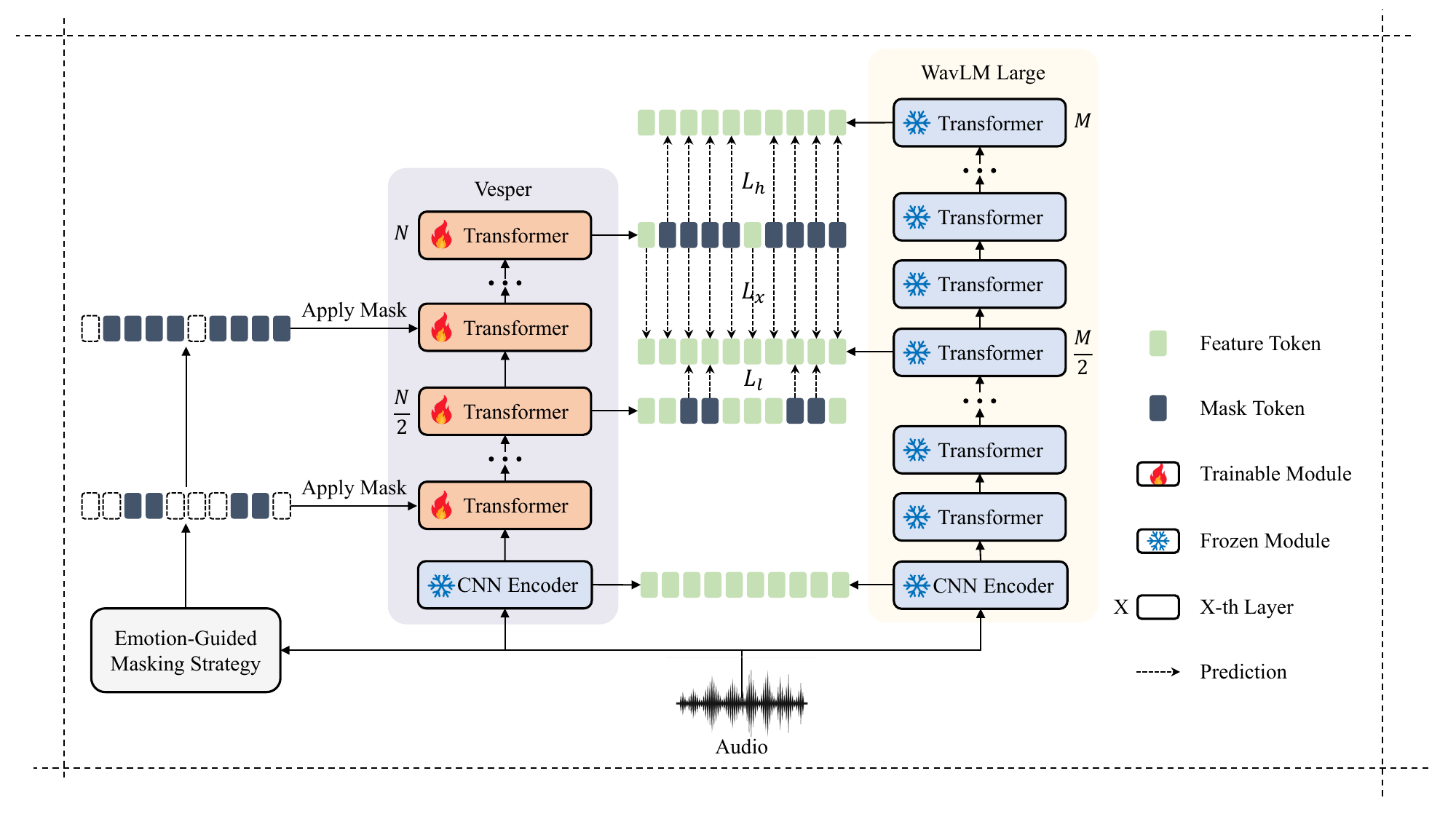}
\caption{The task-specific self-supervised pretraining strategy of the proposed Vesper, which mainly consists of emotion-guided masking strategy, hierarchical self-supervision ($L_l$ and $L_h$), and cross-layer self-supervision ($L_x$). Raw audio samples are used as inputs. $N$ and $M$ ($N < M$) denote the number of Transformer layers employed in Vesper and WavLM Large, respectively.}
\label{Vesper}
\end{figure*}

\subsection{Task-Specific Pretraining by Self-Supervision}

By employing the initialization method above, we can obtain a compact but general Vesper. To ensure that Vesper becomes emotion-specific, it is essential to employ the emotion-guided masking strategy and further pretrain Vesper by the following self-supervised approaches. The overall pretraining strategy for Vesper is illustrated in Fig.~\ref{Vesper}.

\subsubsection{Emotion-Guided Masking Strategy}

\begin{figure}[t]
\centering
\includegraphics[width=0.85\linewidth]{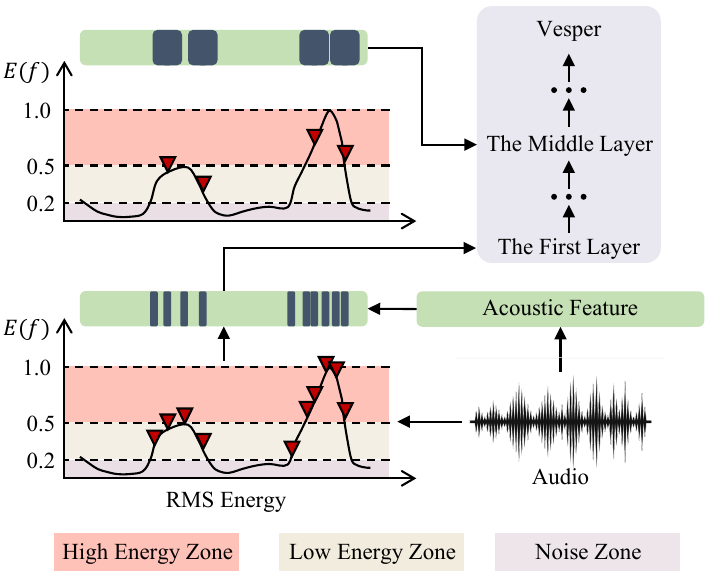}
\caption{The emotion-guided masking strategy used for pretraining Vesper. Only regions with a high probability of containing emotional information are masked. The red triangle indicates the position of the mask center.}
\label{mask}
\end{figure}

To enhance Vesper's sensitivity to emotional information, we propose an emotion-guided masking strategy that is both computationally efficient and effective. In contrast to previous masking strategies that employ random selection of mask positions, we determine the positions to be masked based on the root mean square (rms) energy of the input speech signal. The reason is that the energy of input speech, which represents the loudness and intensity of a speaker's voice, can provide valuable information about the speaker's emotional state. Changes in signal energy indicate variations in emotional states \cite{energy_matter, energy_matter_2, energy_matter_3}. For example, more intense emotional states such as anger or excitement are usually associated with higher signal energy, while more depressed emotional states such as sadness or depression have lower signal energy compared to the normal state. The rms energy is defined as:
\begin{equation}
    E(f) = \sqrt{\frac{1}{L}\sum_{l=1}^L |A_{f}(l)|^2}
    \label{energy}
\end{equation}
where $A$ is the input audio, $A_{f}$ denotes the $f$-th frame of $A$ with frame length $L$ and $E(f)$ denotes the rms energy of the $f$-th frame.

Given the energy $E$, we first divide it by its maximum value to limit the value of $E$ to $[0, 1]$. As depicted in Fig.~\ref{mask}, we then partition the energy into three zones, \textit{i.e.}, a high energy zone with a range of $(0.5, 1]$, a low energy zone with a range of $(0.2, 0.5]$, and a noise zone with a range of $[0, 0.2]$. As mentioned, frames located in the high-energy zone have a high probability of containing intense emotions, while those located in the low-energy zone have a high probability of containing depressed emotions. Therefore, we randomly select half of the mask positions from the high-energy zone and the other half from the low-energy zone to enable Vesper to better focus on emotional information. Eventually, the selected mask position will become the center of the masking region.

During the masking process, we hope to integrate the inherent structural characteristics of speech signals to enhance the interpretability of Vesper. Following SpeechFormer++ \cite{speechformer++}, which initially focuses on phoneme-level modeling followed by word-level modeling, we propose a similar strategy of applying phoneme-level masking first, followed by word-level masking. Specifically, as shown in Fig.~\ref{mask} and the left part of Fig.~\ref{Vesper}, a phoneme-level mask, with a mask span of 160 ms, is first applied to the input of the first Transformer layer in Vesper to promote fine-grained information learning. An additional word-level mask, with a mask span of 800 ms, is applied to the middle Transformer layer to enhance coarse-grained feature learning. The decision to use mask spans of 160 ms and 800 ms is based on the statistical analysis of phoneme and word durations conducted in \cite{speechformer++}. The word-level mask inherits the masking positions from the phoneme-level mask. However, we decrease the number of masking positions in the word-level mask due to its larger mask span.
The masking strategy functions as an auxiliary component, filtering out non-essential regions to facilitate the following self-supervised learning.

\subsubsection{Hierarchical Self-Supervision}

To incorporate emotional specificity into Vesper, we must determine how emotion is expressed in the speech signal. Note that acoustic features such as rhythm and articulation convey many messages about emotion. Meanwhile, semantic information contained in speech reveals the emotional state of the speaker. Therefore, there is an urgent need to enhance Vesper's ability to extract both acoustic and semantic information for emotion recognition. Recent research \cite{12layer} has reported that the shallow layers of pretrained models tend to learn acoustic features, while the deep layers tend to extract the semantic features of speech signals. Accordingly, as shown in Fig.~\ref{Vesper}, we propose the hierarchical self-supervision approach to supervise the shallow and deep layers of Vesper separately. The shallow and deep layers of the frozen WavLM Large model are utilized to generate the targets during the pretraining stage of Vesper.

Specifically, suppose the $j$-th Transformer layer in WavLM Large is ${Tr}_j^W$. The calculation flow in WavLM Large is:
\begin{equation}
    y_0 = CNN(A)
\end{equation}
\begin{equation}
    y_j = {Tr}_j^W(y_{j-1})
\end{equation}
where $j \in [1, M]$; the CNN encoder consumes audio $A$ to yield the latent representation $y_0 \in \mathbb{R}^{T \times d}$, and $T$ and $d$ are the sequence length and dimension, respectively.

Suppose the $i$-th Transformer layer in Vesper is ${Tr}_i^V$, the masked indices in the phoneme-level mask are $I_p$, and the masked indices in the word-level mask are $I_w$. We assume the mask embedding to be $MK \in \mathbb{R}^d$. The calculation flow in Vesper is as follows:
\begin{equation}
    x_0 = CNN(A)
\end{equation}
\begin{equation}
    x^{\prime}_0 = Add\_Mask(x_0, MK, I_p)
\end{equation}
\begin{equation}
    x^{\prime}_i = {Tr}_i^V(x^{\prime}_{i-1}), \quad i \in [1, \frac{N}{2}]
\end{equation}
\begin{equation}
    x^{\prime\prime}_{\frac{N}{2}} = Add\_Mask(x^{\prime}_{\frac{N}{2}}, MK, I_w)
\end{equation}
\begin{equation}
    x^{\prime\prime}_i = {Tr}_i^V(x^{\prime\prime}_{i-1}), \quad i \in [\frac{N}{2}+1, N]
\end{equation}
where $i \in [1, N]$, $Add\_Mask(x, MK, I)$ replaces the embedding in $x$ with $MK$ according to the position index $I$. 

During pretraining, the intermediate and final outputs of Vesper are required to predict the intermediate and final outputs of the WavLM Large model, respectively. The mean squared error (MSE) is utilized to calculate the training loss on the masked regions. Formally, the loss is defined as:
\begin{equation}
    L_l = \sum_{m \in I_p} MSE(\mathcal{P}_1({Tr}_\frac{N}{2}^V(x^{\prime}_{\frac{N}{2}-1}))_m, {Tr}_\frac{M}{2}^W(y_{\frac{M}{2}-1})_m)
\end{equation}
\begin{equation}
    L_h = \sum_{m \in I_w} MSE(\mathcal{P}_2({Tr}_N^V(x^{\prime\prime}_{N-1}))_m, {Tr}_M^W(y_{M-1})_m)
\end{equation}
where $\mathcal{P}_1$ and $\mathcal{P}_2$ are predictors consisting of two linear layers with an activation layer in between.

\subsubsection{Cross-Layer Self-Supervision}

Although hierarchical self-supervision enhances the extraction of both acoustic and semantic information, information bias still persists. The shallow layers of the model excel at extracting acoustic information, while the deep layers are more adept at capturing semantic information. This information bias leads to the representations of different layers in Vesper being somewhat complementary. To make the final representation of Vesper more informative, \textit{i.e.}, containing both acoustic and semantic information, we propose the cross-layer self-supervision approach to provide additional supervision for the learning process of the last layer in Vesper. Formally, the additional cross-layer self-supervision loss is given by
\begin{equation}
    L_x = \sum_{m} MSE(\mathcal{P}_3({Tr}_N^V(x^{\prime\prime}_{N-1}))_m, {Tr}_\frac{M}{2}^W(y_{\frac{M}{2}-1})_m)
\end{equation}
where $\mathcal{P}_3$ is a predictor consuming the final output of Vesper to predict the intermediate output of WavLM Large. Note that $L_x$ is applied to all positions, including both the masked and unmasked parts. The output of Vesper now encompasses both rich semantic information and sufficient acoustic information.

Finally, the objective for training Vesper can be written as:
\begin{equation}
    L = \lambda_l L_l + \lambda_h L_h + \lambda_x L_x
\end{equation}
where $\lambda_l$, $\lambda_h$ and $\lambda_x$ are hyperparameters employed to balance different loss components.

\section{Experimental Setup}

\subsection{Datasets}
Vesper is first pretrained on the LSSED dataset and then fine-tuned on the IEMOCAP, MELD, and CREMA-D datasets to evaluate the performance of speech emotion recognition.

LSSED \cite{lssed} is a recently released large-scale English speech emotion dataset. It comprises data collected from 820 subjects and consists of 147,025 samples. The average duration of all samples in the LSSED is 5.05 s, and the total duration is approximately 206 hours. We do not utilize the sentiment labels provided in the LSSED dataset, as Vesper is pretrained in a self-supervised manner.

IEMOCAP \cite{IEMOCAP} dataset is widely used in the field of speech emotion recognition. It comprises 12 hours of audio data divided into five sessions, each containing one male and one female speaker. In this study, we focus on 5,531 utterances from four emotion categories: angry, neutral, happy\footnote{We merge the excited samples with the happy samples in IEMOCAP.}, and sad. We employ the speaker-independent 5-fold cross-validation strategy to evaluate model performance. The reported results are the average scores of the 5-fold experiments.

MELD \cite{meld} consists of 13,708 utterances extracted from the \textit{Friends} TV series and is categorized into seven emotion classes: anger, disgust, sadness, joy, neutral, surprise, and fear. MELD is officially split into training, validation, and testing sets. We utilize the validation set for hyperparameter tuning. The model with the best performance on the validation set is then evaluated on the testing set. The reported results are the scores achieved on the testing set.

CREMA-D \cite{crema} is an audiovisual dataset with 7442 original clips from 91 actors (48 males and 43 females) between the ages of 20 and 74 across diverse races and ethnicities. 
The actors deliver their lines from a predefined set of 12 sentences, encompassing six different emotional states (anger, disgust, fear, happiness, neutral, and sadness) across four emotional levels (low, medium, high, and unspecified). The model is tested using the standard rule of 80/20, where 80\% of the samples are used for training and 20\% are used for testing. The scores achieved on the testing set are reported.

\subsection{Evaluation Metrics}

We apply three widely used evaluation metrics to evaluate the performance of speech emotion recognition: weighted accuracy (WA), unweighted accuracy (UA), and weighted average F1 (WF1). The metrics are computed as follows:
\begin{align}
    WA &= \frac{1}{\sum_{c=1}^CN_c}\sum_{c=1}^C N_c\times Acc(c)\\
    UA &= \frac{1}{C}\sum_{c=1}^C Acc(c)\\
    WF1 &= \frac{1}{\sum_{c=1}^CN_c}\sum_{c=1}^C N_c\times F1(c)
    \label{Metric}
\end{align}
where $N_c$ denotes the number of samples of the $c$-th category and $Acc(c)$ and $F1(c)$ are the classification accuracy and F1 score of the $c$-th category, respectively.
We adopt the WA as our primary metric to evaluate the IEMOCAP and CREMA-D datasets. 
For the MELD dataset, we select WF1 as our primary evaluation metric because of the sample imbalance issue.

\subsection{Implementation Details}

\begin{table}[t]
    \caption{Hyperparameters for pretraining and fine-tuning Vesper}
    \label{tab_hyper}
    \centering
    \begin{threeparttable}
    \begin{tabular}{c|cc}
    \hline
    Hyperparameters     &  Pretraining  & Fine-tuning  \\ \hline
    Training epochs     &  100           & 50    \\
    Warmup epochs       &  5             & -     \\
    Batch size          &  32            & 32    \\
    Base learning rate       &  5e-4          & 7e-4  \\
    Optimizer           &  AdamW \cite{adamw}        & SGD \cite{sgd}    \\ 
    Optimizer momentum  & $\beta_1, \beta_2 = 0.9, 0.999$  & 0.9  \\
    Weight decay        &  0.01          & 0.01   \\
    Learning rate schedule    &  Cosine annealing \cite{cos_decay} & Cosine annealing \\
    Minimum learning rate & 5e-6       & 7e-6  \\ \hline
    \end{tabular}
    \end{threeparttable}
\end{table}

When fine-tuning on the downstream datasets, the cross-entropy loss is employed as the objective function. Unless otherwise stated, the downstream model is implemented as a simple classifier consisting of two fully connected layers with an average pooling layer in between, which is consistent with the SUPERB benchmark \cite{superb}. The classifier is placed on top of the pretrained Vesper to predict the emotion state. The hidden dimension of the classifier is set to 256. Note that the pretrained Vesper remains fixed, and only the classifier is trained during the fine-tuning process. In addition, we freeze the CNN encoder throughout the process, including during pretraining and fine-tuning. To align with the SUPERB benchmark, the representations of each layer are weighted by trainable weights to generate the input of the downstream classifier, unless otherwise stated. The audio samples in LSSED are cropped or padded to 5s for pretraining. The audio samples in IEMOCAP, MELD, and CREMA-D are cropped or padded to 6.5s, 4.5s, and 3s for fine-tuning, respectively.

We introduce two versions, Vesper-4 and Vesper-12, with 4 and 12 Transformer layers, respectively. Note that the number of Transformer layers employed in WavLM Base is 12, and the number of Transformer layers used in WavLM Large is 24. Due to the shared model structure between Vesper and WavLM, the configurations of Vesper-4 and Vesper-12 remain consistent with WavLM Large, except for the variation in the number of layers employed.
The number of masking positions is set to 20 for the phoneme-level mask and 4 for the word-level mask.
$\lambda_l$, $\lambda_h$ and $\lambda_x$ are empirically set to 1.0, 0.1 and 1.0, respectively. More detailed training hyperparameters are shown in Table~\ref{tab_hyper}. The platform used for model training and testing is an Ubuntu 18.04 server equipped with GeForce RTX 2080 Ti GPU. The program is implemented based on Pytorch.

\section{Results and Discussion}
\label{Results_and_Discussion}

\begin{table}[t]
    \caption{Computational efficiency of WavLM and Vesper on the IEMOCAP, MELD, and CREMA-D datasets. The last three columns indicate the theoretical computational complexity (FLOPs) on the respective datasets} 
    \label{tab_complex}
    \centering
    \begin{threeparttable}
    \begin{tabular}{c|c|ccc}
    \hline
    Method      & \# Params &  IEMOCAP & MELD   & CREMA-D \\ \hline
    WavLM Base  & 94.70M    &  45.15G  & 31.23G & 20.79G  \\
    Vesper-4    & \textbf{63.52M}    &  \textbf{35.16G}  & \textbf{24.33G} & \textbf{16.20G}  \\ \hline
    WavLM Large & 316.62M   &  116.75G & 80.74G & 53.72G  \\
    Vesper-12   & \textbf{164.29M}   &  \textbf{67.80G}  & \textbf{46.89G} & \textbf{31.21G}  \\ \hline
    \end{tabular}
    \end{threeparttable}
\end{table}

\begin{table}[t]
    \caption{Performance of WavLM and Vesper on the IEMOCAP, MELD, and CREMA-D datasets for speech emotion recognition}
    \label{tab_ser}
    \centering
    \begin{threeparttable}
    \begin{tabular}{c|c|ccc}
    \hline
    Dataset & Method & WA & UA & WF1  \\ \hline
    \multirow{4}{*}{IEMOCAP} & WavLM Base  & 0.659\tnote{$\dagger$} & -  & -  \\
    & Vesper-4    & \textbf{0.684} & \textbf{0.693} & \textbf{0.683}   \\ \cline{2-5}
    & WavLM Large & 0.706\tnote{$\dagger$} & - & - \\
    & Vesper-12   & \textbf{0.707} & \textbf{0.708} & \textbf{0.706}  \\ \hline
    \multirow{4}{*}{MELD} & WavLM Base & 0.499 & 0.201 & 0.400   \\
    & Vesper-4    & \textbf{0.501} & \textbf{0.250} & \textbf{0.457}   \\ \cline{2-5}
    & WavLM Large & \textbf{0.542} & 0.253 & 0.476  \\
    & Vesper-12   & 0.535 & \textbf{0.268} & \textbf{0.480}  \\ \hline
    \multirow{4}{*}{CREMA-D} & WavLM Base & 0.599 & 0.599 & 0.600   \\
    & Vesper-4    & \textbf{0.734} & \textbf{0.737} & \textbf{0.733}   \\ \cline{2-5}
    & WavLM Large & 0.757 & 0.762 & 0.755  \\
    & Vesper-12   & \textbf{0.772} & \textbf{0.776} & \textbf{0.768}  \\ \hline
    \end{tabular}
    \begin{tablenotes}
        \footnotesize
        \item[$\dagger$] Results from the official WavLM paper \cite{wavlm}.
    \end{tablenotes}
    \end{threeparttable}
\end{table}

\subsection{Performance and Computational Efficiency}

We compare the proposed Vesper-4 with the general pretrained WavLM Base and compare Vesper-12 with WavLM Large in terms of performance and computational efficiency. Specifically, the number of parameters and the theoretical computational complexity (FLOPs) are listed in Table~\ref{tab_complex}, and the performance of speech emotion recognition is reported in Table~\ref{tab_ser}. Compared to the WavLM Base model, Vesper-4 has a smaller model size (63.52 M vs. 94.70 M, relative reduction of 32.9\%) but is able to deliver superior performances in all metrics. Specifically, our Vesper-4 results in a 2.5\% WA gain over WavLM Base on the IEMOCAP dataset and 13.5\% WA gain over WavLM Base on the CREMA-D dataset. On the MELD dataset, Vesper-4 improves the accuracies (0.2\% in WA, 4.9\% in UA and 5.7\% in WF1) compared to WavLM Base. Similarly, comparing our Vesper-12 with the WavLM Large model shows that the model size of Vesper-12 is approximately half that of the WavLM Large model (164.29 M vs. 316.62 M, relative reduction of 48.1\%). However, Vesper-12 achieves comparable results to WavLM Large (0.707 vs. 0.706 in WA on the IEMOCAP dataset, 0.480 vs. 0.476 in WF1 on the MELD dataset, and 0.772 vs. 0.757 in WA on the CREMA-D dataset). In addition, Vesper-4 demonstrates a 22.1\% reduction in computational burden compared to WavLM Base, and the computational burden of Vesper-12 is reduced by 41.9\% compared to WavLM Large. The experimental results indicate that there is indeed room for improvement in using a general pretrained model for specific downstream tasks. Through task-specific continuous pretraining, the general pretrained WavLM model is transformed into Vesper, which is both compact and effective for speech emotion recognition.

\subsection{Performance of Downstream Models with Vesper's Features}
\label{downstream}

\begin{table*}
    \caption{Evaluating various downstream models on the IEMOCAP, MELD, and CREMA-D datasets using diverse upstream features.}
    \label{tab_fea}
    \centering
    \begin{threeparttable}
    \begin{tabular}{c|c|c|ccc|c|ccc}
    \hline
    Dataset & Upstream & Downstream & WA & UA & WF1 & Downstream &  WA & UA & WF1 \\ \hline
    \multirow{4}{*}{IEMOCAP} & WavLM Base & \multirow{12}{*}{Shiftformer \cite{shiftser}} & 0.615 & 0.624 & 0.609 & \multirow{12}{*}{SpeechFormer-S \cite{speechformer}} & 0.608 & 0.613 & 0.602\\
    & Vesper-4 & & \textbf{0.730} & \textbf{0.739} & \textbf{0.730} & & \textbf{0.711} & \textbf{0.725} & \textbf{0.711}  \\ \cline{4-6} \cline{8-10} \cline{2-2}
    & WavLM Large & & 0.727 & 0.736 & 0.727 & & 0.721 & 0.729 & 0.719 \\
    & Vesper-12 & & \textbf{0.737} & \textbf{0.743} & \textbf{0.735} & & \textbf{0.729} & \textbf{0.731} & \textbf{0.725} \\ \cline{1-2} \cline{4-6} \cline{8-10} 
    \multirow{4}{*}{MELD} & WavLM Base & & \textbf{0.480} & 0.212 & 0.424 & & 0.481 & 0.208 & 0.414 \\
    & Vesper-4 & & 0.479 & \textbf{0.240} & \textbf{0.445} & & \textbf{0.489} & \textbf{0.260} & \textbf{0.441} \\ \cline{4-6} \cline{8-10} \cline{2-2}
    & WavLM Large & & 0.527 & \textbf{0.265} & 0.476 & & 0.493 & \textbf{0.281} & 0.474 \\
    & Vesper-12 & & \textbf{0.530} & 0.262 & \textbf{0.479} & & \textbf{0.507} & 0.276 & \textbf{0.477} \\ \cline{1-2} \cline{4-6} \cline{8-10} 
    \multirow{4}{*}{CREMA-D} & WavLM Base & & 0.709 & 0.712 & 0.708 & & 0.696 & 0.698 & 0.691 \\
    & Vesper-4 & & \textbf{0.780} & \textbf{0.783} & \textbf{0.778} & & \textbf{0.761} & \textbf{0.764} & \textbf{0.760} \\ \cline{4-6} \cline{8-10} \cline{2-2}
    & WavLM Large & & 0.751 & 0.754 & 0.751 & & 0.770 & 0.772 & 0.769 \\
    & Vesper-12 & & \textbf{0.806} & \textbf{0.809} & \textbf{0.804} & & \textbf{0.794} & \textbf{0.797} & \textbf{0.795} \\ \hline
    \end{tabular}
    \end{threeparttable}
\end{table*}

We replace the simple classifier used in the SUPERB benchmark with two state-of-the-art approaches, namely, Shiftformer \cite{shiftser} and SpeechFormer \cite{speechformer}, to further evaluate the effectiveness of Vesper. The experimental results are shown in Table~\ref{tab_fea}. When employing the downstream model with Vesper-4 features on the IEMOCAP dataset, we observe an improvement of approximately 10\% across all metrics compared to WavLM Base. Compared to WavLM Large, utilizing Vesper-12 on the IEMOCAP dataset results in an improvement of 1.0\% in WA for Shiftformer and 0.8\% for SpeechFormer. For the MELD dataset, Vesper-4 outperforms WavLM Base by 2.1\% in WF1 for Shiftformer and 2.7\% in WF1 for SpeechFormer. Compared to WavLM Large, the utilization of Vesper-12 on MELD leads to an improvement of 0.3\% in WF1 for both Shiftformer and SpeechFormer. On the CREMA-D dataset, significant improvements are observed when utilizing Vesper features. Specifically, when comparing Vesper-4 to WavLM Base, we observe improvements of 7.1\% and 6.5\% in WA for Shiftformer and SpeechFormer, respectively. Finally, Shiftformer using Vesper-12 achieves a classification accuracy of 80.6\%. We conclude that Vesper yields improved performance across various downstream models.

\subsection{Comparison with Some Known Systems}

We compare our top-performing model (Vesper-12) against previous works on IEMOCAP, MELD, and CREMA-D datasets and the results are presented in Table~\ref{tab_sota}. It is important to note that the prior systems compared with Vesper are specifically tailored for the speech emotion recognition task and demand significant manual tuning, whereas our Vesper, followed by two fully connected layers (FC), is simply and directly applied in the target datasets. As seen in Table~\ref{tab_sota}, our Vesper with FC is able to attain comparable performance against existing methods on the IEMOCAP and MELD datasets. More concretely, our Vesper outperforms MCFN \cite{MCFN}, ISNet \cite{ISNet}, and DAAE \cite{DAAE} systems and achieves comparable results to CA-MSER \cite{CA-MSER} on IEMOCAP. In the MELD dataset, Vesper's performance demonstrates only a marginal difference compared to SpeechFormer++ \cite{speechformer++} in the UA metric. However, across all other metrics, Vesper with FC consistently outperforms its competitors. For the CREMA-D dataset, Vesper with FC outperforms the existing methods by a substantial margin in all the listed metrics. Particularly, the enhancements observed in WA, UA, and WF1 are 2.5\%, 18.9\%, and 19.1\% respectively. The above results once again confirm the superiority and robustness of Vesper. 
Additionally, when Vesper is augmented with the advanced downstream model Shiftformer \cite{shiftser}, significant improvements in the results are observed on the IEMOCAP and CREMA-D datasets. More details and analysis can be found in Section~\ref{downstream}.

\begin{table}[t]
    \caption{Comparison with some known systems on the IEMOCAP, MELD, and CREMA-D datasets. All systems apply audio as input}
    \label{tab_sota}
    \centering
    \begin{threeparttable}
    \begin{tabular}{c|c|ccc}
    \hline
    Dataset & Method & WA & UA & WF1  \\ \hline
    \multirow{6}{*}{IEMOCAP} & MCFN \cite{MCFN}  & 0.621 & 0.603  & -  \\
    & \tnote{$\dagger$}\; ISNet \cite{ISNet} & 0.704 & 0.650 & - \\
    & DAAE \cite{DAAE} & 0.701 & 0.707 & - \\
    & CA-MSER \cite{CA-MSER} & 0.698 & 0.711 & - \\ \cline{2-5}
    & Vesper (FC)  & 0.707 & 0.708 & 0.706  \\
    & Vesper (Shiftformer)  & \textbf{0.737} & \textbf{0.743} & \textbf{0.735}  \\ \hline
    \multirow{6}{*}{MELD} & MCFN \cite{MCFN} & 0.481 & - & 0.367   \\
    & \tnote{$\dagger$}\; \tnote{$\ddagger$}\; DECN \cite{DECN} & 0.493 & - & 0.439 \\
    & \tnote{$\dagger$}\; \tnote{$\ddagger$}\; SCFA \cite{SCFA} & 0.474  & - & 0.442 \\
    & SpeechFormer++ \cite{speechformer++} & 0.510 & \textbf{0.273} & 0.470 \\ \cline{2-5}
    & Vesper (FC)   & \textbf{0.535} & 0.268 & \textbf{0.480}  \\
    & Vesper (Shiftformer)  & 0.530 & 0.262 & 0.479  \\ \hline
    \multirow{6}{*}{CREMA-D} & IG-CNN \cite{IG-CNN} & 0.580 & 0.585 & 0.577 \\
    & LanSER \cite{LanSER} & - & 0.587 & - \\
    & Phukan \cite{Phukan} & 0.705 & - & -   \\
    & fusion\_cat\_xwc \cite{wu} & 0.747 & - & - \\ \cline{2-5}
    & Vesper (FC)   & 0.772 & 0.776 & 0.768  \\
    & Vesper (Shiftformer)  & \textbf{0.806} & \textbf{0.809} & \textbf{0.804}  \\ \hline
    \end{tabular}
    \begin{tablenotes}
        \footnotesize
        \item[$\dagger$] Speaker information is used.
        \item[$\ddagger$] Conversation context is used.
    \end{tablenotes}
    \end{threeparttable}
\end{table}

\subsection{Ablation Study}
In this section, we conduct a comprehensive ablation study to demonstrate the effectiveness of the various components of Vesper.
The downstream classifier utilized in the following experiments is consistent with the SUPERB benchmark.

\subsubsection{Comparison of Different Compression Methods}

\begin{table}[t]
    \caption{Performance of Vesper using different compression methods on the IEMOCAP, MELD, and CREMA-D datasets}
    \label{tab_init}
    \centering
    \begin{threeparttable}
    \begin{tabular}{c|c|c|ccc}
    \hline
    Method & Dataset & Compression  & WA  & UA  & WF1   \\ \hline
    \multirow{12}{*}{Vesper-4} & \multirow{4}{*}{IEMOCAP} & Random & 0.534 & 0.551 & 0.524 \\
    & & Distillation  & 0.665 & 0.670 & 0.667  \\
    & & Averaging  & 0.659 & 0.663 & 0.654  \\
    & & Extraction (used) & \textbf{0.684} & \textbf{0.693} & \textbf{0.683}  \\  \cline{2-6}
    & \multirow{4}{*}{MELD} & Random & 0.438 & 0.162 & 0.291  \\
    & & Distillation  & 0.499 & 0.225 & 0.437  \\
    & & Averaging  & 0.488 & \textbf{0.261} & 0.440  \\
    & & Extraction (used) & \textbf{0.501} & 0.250 & \textbf{0.457}  \\  \cline{2-6}
    & \multirow{4}{*}{CREMA-D} & Random & 0.595 & 0.616 & 0.596  \\
    & & Distillation  & 0.712 & 0.722 & 0.714  \\
    & & Averaging  & 0.715 & 0.707 & 0.709  \\
    & & Extraction (used) & \textbf{0.734} & \textbf{0.737} & \textbf{0.733}  \\  \hline
    \multirow{12}{*}{Vesper-12} & \multirow{4}{*}{IEMOCAP} & Random & 0.552 & 0.562 & 0.547 \\
    & & Distillation  & 0.690 & 0.700 & 0.687  \\
    & & Averaging  & 0.687 & 0.697 & 0.679  \\
    & & Extraction (used) & \textbf{0.707} & \textbf{0.708} & \textbf{0.706}  \\  \cline{2-6}
    & \multirow{4}{*}{MELD} & Random & 0.484 & 0.154 & 0.336  \\
    & & Distillation  & 0.516 & 0.255 & 0.463  \\
    & & Averaging  & 0.512 & \textbf{0.277} & 0.461  \\
    & & Extraction (used) & \textbf{0.535} & 0.268 & \textbf{0.480}  \\  \cline{2-6}
    & \multirow{4}{*}{CREMA-D} & Random & 0.620 & 0.623 & 0.616  \\
    & & Distillation  & 0.766 & 0.761 & 0.762  \\
    & & Averaging  & 0.759 & 0.760 & 0.761  \\
    & & Extraction (used) & \textbf{0.772} & \textbf{0.776} & \textbf{0.768}  \\  \hline
    \end{tabular}
    \end{threeparttable}
\end{table}

Taking advantage of the common dimensionality between WavLM Large and Vesper networks, we directly employ the parameters of WavLM Large to initialize Vesper and simultaneously achieve compression. Specifically, direct compression includes uniform extraction and uniform averaging. To probe the effect of different compression methods, we compare direct compression with knowledge distillation. The results are shown in Table~\ref{tab_init}. All compression methods are described in Section~\ref{sec_init}. We also implement random initialization, which assigns random initial parameters to Vesper, for a comprehensive comparison. The pretraining process is consistent for all experiments. As shown in Table~\ref{tab_init}, random initialization leads to the worst performance in all cases due to the lack of training data. Thus, considering the knowledge acquired from the pretrained WavLM is essential. Distillation-based compression leverages universal knowledge and obtains 0.665 WA on IEMOCAP, 0.437 WF1 on MELD and 0.712 on CREMA-D for Vesper-4. For Vesper-12, distillation compression yields 0.690 WA on IEMOCAP, 0.463 WF1 on MELD, and 0.766 WA on CREMA-D. Nevertheless, the uniform averaging method achieves comparable performance to distillation with simpler operations on the three datasets, exhibiting a difference in evaluation metrics of less than 1\%. Uniform extraction, which directly copies the parameters of WavLM Large to Vesper, further simplifies the compression process and achieves the best performance in all cases. The results indicate that directly initializing Vesper with the parameters of the pretrained model is not only the simplest approach but also the most effective. Hence, we apply uniform extraction as the compression and initialization method by default.

\subsubsection{Effectiveness of the Emotion-Guided Masking Strategy} 

\begin{table}[t]
    \caption{Performance of Vesper using a different masking strategy on the IEMOCAP, MELD, and CREMA-D datasets}
    \label{tab_mask}
    \centering
    \begin{threeparttable}
    \begin{tabular}{c|c|c|ccc}
    \hline
    Method & Dataset & Masking  & WA  & UA  & WF1   \\ \hline
    \multirow{6}{*}{Vesper-4} & \multirow{2}{*}{IEMOCAP} & Random & 0.656 & 0.667 & 0.662 \\
    & & Emotion-Guided & \textbf{0.684} & \textbf{0.693} & \textbf{0.683} \\ \cline{2-6}
    & \multirow{2}{*}{MELD} & Random & \textbf{0.506} & 0.227 & 0.445  \\
    & & Emotion-Guided & 0.501 & \textbf{0.250} & \textbf{0.457} \\ \cline{2-6}
    & \multirow{2}{*}{CREMA-D} & Random & 0.718 & 0.721 & 0.717  \\
    & & Emotion-Guided & \textbf{0.734} & \textbf{0.737} & \textbf{0.733} \\ \hline
    \multirow{6}{*}{Vesper-12} & \multirow{2}{*}{IEMOCAP} & Random & 0.688 & 0.691& 0.685 \\
    & & Emotion-Guided & \textbf{0.707} & \textbf{0.708} & \textbf{0.706} \\ \cline{2-6}
    & \multirow{2}{*}{MELD} & Random & 0.517 & 0.260 & 0.469  \\
    & & Emotion-Guided & \textbf{0.535} & \textbf{0.268} & \textbf{0.480} \\ \cline{2-6}
    & \multirow{2}{*}{CREMA-D} & Random & 0.762 & 0.756 & 0.760  \\
    & & Emotion-Guided & \textbf{0.772} & \textbf{0.776} & \textbf{0.768} \\ \hline
    \end{tabular}
    \end{threeparttable}
\end{table}

In contrast to the traditional masking strategy, which randomly determines the masking position, the proposed emotion-guided masking strategy enhances Vesper's sensitivity to emotional information by determining the masking position according to the rms energy. Table~\ref{tab_mask} compares these two masking strategies. For Vesper-4 on the IEMOCAP dataset, the emotion-guided masking strategy shows an absolute improvement of 2.8\% in WA, 2.6\% in UA, and 2.1\% in WF1 over the traditional random masking strategy. For the MELD dataset, although the random strategy obtains a slightly higher WA score (0.506 vs. 0.501 in WA), the primary evaluation metric WF1 decreases by 1.2\% compared to the emotion-guided masking strategy. For the CREMA-D dataset, the proposed strategy outperforms random masking by 1.6\% on all evaluation metrics. In the case of Vesper-12, the proposed masking strategy shows an absolute improvement of 1.9\% in WA, 1.7\% in UA, and 2.1\% in WF1 over the random strategy on the IEMOCAP dataset. For the MELD dataset, Vesper-12 equipped with the emotion-guided masking strategy achieves a gain of 1.8\% in WA, 0.8\% in UA, and 1.1\% in WF1. For the CREMA-D dataset, our strategy outperforms random masking by 1.0\% on WA, 2.0\% on UA, and 0.8\% on WF1. These results verify the effectiveness of the emotion-guided masking strategy for speech emotion recognition.

\subsubsection{Effectiveness of Hierarchical Self-Supervision}

To verify the potency of the hierarchical self-supervision, we conduct experiments by discarding the loss $L_l$ or solely applying the loss $L_h$. The experimental results are shown in Table~\ref{tab_loss}. It can be seen that the absence of $L_l$ leads to decreased performance on all datasets. For example, Vesper-4 exhibits a decrease of 2.3\% in WA on the IEMOCAP dataset, a decrease of 0.6\% in WF1 on the MELD dataset, and a decrease of 2.0\% in WA on the CREMA-D dataset when $L_l$ is omitted. A similar trend is observed in Vesper-12 when $L_l$ is omitted, which results in an absolute reduction of 1.6\% in WA on IEMOCAP, 1.0\% in WF1 on MELD, and 1.0\% in WA on the CREMA-D. Self-supervision of the lower layers is completely eliminated when applying only the $L_h$ loss. However, the results presented in Table~\ref{tab_loss} demonstrate that relying solely on the $L_h$ loss yields the poorest performance across all datasets, with an absolute reduction of 1.0\%$\sim$3.7\% in all evaluation metrics. These results confirm the effectiveness of the proposed hierarchical self-supervision approach for speech emotion recognition.

\begin{table}[t]
    \caption{Performance of Vesper using different combinations of losses on the IEMOCAP, MELD, and CREMA-D datasets}
    \label{tab_loss}
    \centering
    \begin{threeparttable}
    \begin{tabular}{c|c|ccc|ccc}
    \hline
    Method & Dataset & $L_l$  & $L_h$  & $L_x$ &  WA & UA & WF1   \\ \hline
    \multirow{12}{*}{Vesper-4} & \multirow{4}{*}{IEMOCAP} & & $\surd$ & & 0.647 & 0.662 & 0.643  \\
    & & $\surd$ & $\surd$ &        & 0.668 & 0.682 & 0.663  \\
    & & & $\surd$ & $\surd$        & 0.661 & 0.673 & 0.658  \\
    & & $\surd$ & $\surd$ & $\surd$ & \textbf{0.684} & \textbf{0.693} & \textbf{0.683}  \\ \cline{2-8}
    & \multirow{4}{*}{MELD} &  & $\surd$ & & 0.506 & 0.233 & 0.447 \\
    & & $\surd$ & $\surd$ &        & 0.499 & 0.249 & 0.452  \\
    & & & $\surd$ & $\surd$        & \textbf{0.508} & 0.240 & 0.451  \\
    & & $\surd$ & $\surd$ & $\surd$ & 0.501 & \textbf{0.250} & \textbf{0.457}  \\ \cline{2-8}
    & \multirow{4}{*}{CREMA-D} &  & $\surd$ & & 0.702 & 0.705 & 0.702 \\
    & & $\surd$ & $\surd$ &        & 0.723 & 0.727 & 0.722  \\
    & & & $\surd$ & $\surd$        & 0.714 & 0.718 & 0.712  \\
    & & $\surd$ & $\surd$ & $\surd$ & \textbf{0.734} & \textbf{0.737} & \textbf{0.733}  \\ \hline
    \multirow{12}{*}{Vesper-12} & \multirow{4}{*}{IEMOCAP} & & $\surd$ & & 0.682 & 0.700 & 0.678  \\
    & & $\surd$ & $\surd$ &        & 0.699 & 0.704 & 0.701  \\
    & & & $\surd$ & $\surd$        & 0.691 & 0.698 & 0.691  \\
    & & $\surd$ & $\surd$ & $\surd$ & \textbf{0.707} & \textbf{0.708} & \textbf{0.706}  \\ \cline{2-8}
    & \multirow{4}{*}{MELD} &  & $\surd$ & & 0.515 & 0.261 & 0.469  \\
    & & $\surd$ & $\surd$ &        & 0.523 & 0.249 & 0.464  \\
    & & & $\surd$ & $\surd$        & 0.522 & 0.256 & 0.470  \\
    & & $\surd$ & $\surd$ & $\surd$ & \textbf{0.535} & \textbf{0.268} & \textbf{0.480}  \\ \cline{2-8}
    & \multirow{4}{*}{CREMA-D} &  & $\surd$ & & 0.755 & 0.759 & 0.751 \\
    & & $\surd$ & $\surd$ &        & 0.766 & 0.760 & 0.766  \\
    & & & $\surd$ & $\surd$        & 0.762 & 0.766 & 0.760  \\
    & & $\surd$ & $\surd$ & $\surd$ & \textbf{0.772} & \textbf{0.776} & \textbf{0.768}  \\ \hline
    \end{tabular}
    \end{threeparttable}
\end{table}

\subsubsection{Effectiveness of Cross-Layer Self-Supervision}

To validate the indispensability of the cross-layer self-supervision, we discard the loss $L_x$ to invalidate it. The results are summarized in Table~\ref{tab_loss}. Taking Vesper-4 as an example, on the IEMOCAP dataset, when the loss $L_x$ is disabled, WA decreases from 0.684 to 0.668, and UA decreases from 0.693 to 0.682. On the MELD dataset, WF1 decreases from 0.457 to 0.452 under the same conditions. For the CREMA-D dataset, disabling the loss $L_x$ leads to a decrease of 0.8\% in WA, 1.0\% in UA, and 1.1\% in WF1. The performance of Vesper-12 is also weakened on all datasets when cross-layer self-supervision is disabled. In particular, on the IEMOCAP dataset, omitting the loss $L_x$ results in a decrease of 0.8\% in WA. Similarly, on the MELD dataset, the primary evaluation metric WF1 decreases by 1.6\%. On the CREMA-D dataset, WA decreases by 0.6\% and UA decreases by 1.6\%. These experimental results confirm the necessity of cross-layer self-supervision.

\subsection{Incorporating Pitch Changes in the Masking Strategy}

In the pretraining phase, Vesper has to predict the latent representations of the masked regions, which are selected based on the energy of the speech signal. As shown in \cite{energy_matter}, when coupled with changes in pitch, energy can be useful for emotion recognition. In this subsection, we integrate both energy and pitch changes into the masking strategy, requiring the selected regions to encompass substantial pitch variations. We aim to investigate whether these additional pitch cues enhance the current masking strategy to be more ``emotion-guided". The experimental results are shown in Table~\ref{tab_pitch}. When the masking strategy incorporates information about pitch changes, we can observe that the improvement ranges from 0.2\% to 0.4\% in the majority of cases. These results indicate that incorporating additional emotion-relevant speech features in the masking strategy can be advantageous. This inclusion aids in identifying crucial regions containing emotional information with high probability, thereby enhancing Vesper's sensitivity to emotional cues. In conclusion, this subsection demonstrates the potential of the emotion-guided masking strategy and suggests the possibility of further improvement.

\begin{table}[t]
    \caption{Performance of Vesper with the inclusion of pitch changes in the masking strategy on the IEMOCAP, MELD, and CREMA-D datasets. $\Delta$ denotes an improvement (+) or a reduction (-)}
    \label{tab_pitch}
    \centering
    \begin{threeparttable}
    \begin{tabular}{c|c|c|ccc}
    \hline
    Method & Dataset & Masking & WA  & UA  & WF1   \\ \hline
    \multirow{9}{*}{Vesper-4} & \multirow{3}{*}{IEMOCAP} & Energy & 0.684 & 0.693 & 0.683 \\
    & & Energy + Pitch & 0.686 & 0.696 & 0.688 \\ \cline{3-6}
    & & $\Delta$ & +0.2\% & +0.3\% & +0.5\% \\ \cline{2-6}
    & \multirow{3}{*}{MELD} & Energy & 0.501 & 0.250 & 0.457  \\
    & & Energy + Pitch & 0.502 & 0.254 & 0.460 \\ \cline{3-6}
    & & $\Delta$ & +0.1\% & +0.4\% & +0.3\% \\ \cline{2-6}
    & \multirow{3}{*}{CREMA-D} & Energy & 0.734 & 0.737 & 0.733  \\
    & & Energy + Pitch & 0.737 & 0.739 & 0.737 \\ \cline{3-6}
    & & $\Delta$ & +0.3\% & +0.2\% & +0.4\% \\ \hline
    \multirow{9}{*}{Vesper-12} & \multirow{3}{*}{IEMOCAP} & Energy & 0.707 & 0.708 & 0.706 \\
    & & Energy + Pitch & 0.709 & 0.710 & 0.709 \\ \cline{3-6}
    & & $\Delta$ & +0.2\% & +0.2\% & +0.3\% \\ \cline{2-6}
    & \multirow{3}{*}{MELD} & Energy & 0.535 & 0.268 & 0.480  \\
    & & Energy + Pitch & 0.538 & 0.272 & 0.483 \\ \cline{3-6}
    & & $\Delta$ & +0.3\% & +0.4\% & +0.3\% \\ \cline{2-6}
    & \multirow{3}{*}{CREMA-D} & Energy & 0.772 & 0.776 & 0.768  \\
    & & Energy + Pitch & 0.775 & 0.778 & 0.773 \\ \cline{3-6}
    & & $\Delta$ & +0.4\% & +0.2\% & +0.5\% \\ \hline
    \end{tabular}
    \end{threeparttable}
\end{table}

\subsection{Introducing Extra Supervision Signals from More Layers}

\begin{table}[t]
    \caption{Performance of introducing additional supervisory signals from WavLM's Transformer layers on the IEMOCAP, MELD, and CREMA-D datasets. Target Layers indicates the Transformer layers that provide supervisory signals to Vesper in hierarchical self-supervision}
    \label{tab_add}
    \centering
    \begin{threeparttable}
    \begin{tabular}{c|c|c|ccc}
    \hline
    Method & Dataset & Target Layers & WA  & UA  & WF1   \\ \hline
    \multirow{9}{*}{Vesper-4} & \multirow{3}{*}{IEMOCAP} & 12, 24 & 0.684 & 0.693 & 0.683 \\
    & & 12, 18, 24 & 0.682 & 0.697 & 0.681 \\
    & & 6, 12, 18, 24 & \textbf{0.686} & \textbf{0.699} & \textbf{0.687} \\ \cline{2-6}
    & \multirow{3}{*}{MELD} & 12, 24 & 0.501 & \textbf{0.250} & 0.457  \\
    & & 12, 18, 24 & 0.505 & 0.248 & 0.458 \\
    & & 6, 12, 18, 24 & \textbf{0.507} & 0.249 & \textbf{0.460} \\ \cline{2-6}
    & \multirow{3}{*}{CREMA-D} & 12, 24 & \textbf{0.734} & \textbf{0.737} & \textbf{0.733}  \\
    & & 12, 18, 24 & 0.730 & 0.735 & 0.728 \\
    & & 6, 12, 18, 24 & 0.731 & 0.736 & 0.730 \\ \hline
    \multirow{9}{*}{Vesper-12} & \multirow{3}{*}{IEMOCAP} & 12, 24 & \textbf{0.707} & 0.708 & \textbf{0.706} \\
    & & 12, 18, 24 & 0.704 & 0.706 & 0.698 \\
    & & 6, 12, 18, 24 & 0.702 & \textbf{0.710} & 0.696 \\ \cline{2-6}
    & \multirow{3}{*}{MELD} & 12, 24 & \textbf{0.535} & 0.268 & \textbf{0.480}  \\
    & & 12, 18, 24 & 0.529 & \textbf{0.270} & 0.479 \\
    & & 6, 12, 18, 24 & 0.521 & 0.269 & 0.477 \\ \cline{2-6}
    & \multirow{3}{*}{CREMA-D} & 12, 24 & \textbf{0.772} & \textbf{0.776} & 0.768  \\
    & & 12, 18, 24 & 0.771 & 0.775 & \textbf{0.769} \\
    & & 6, 12, 18, 24 & 0.764 & 0.768 & 0.761 \\ \hline
    \end{tabular}
    \end{threeparttable}
\end{table}

Currently we supervise the intermediate and the final layers of Vesper. 
We are interested in assessing the performance of introducing additional supervision signals from WavLM to guide the learning of other layers in Vesper. Therefore, we use the 18th Transformer layer output from WavLM to supervise the learning of Vesper-4's 3rd layer or Vesper-12's 9th layer. Moreover, we employ the 6th layer output from WavLM to guide the learning of Vesper-4's 1st layer or Vesper-12's 3rd layer. Note that the proposed supervision for the intermediate and final layers is consistently applied. The recognition results on three corpora are reported in Table~\ref{tab_add}. Unexpectedly, the addition of extra supervision signals does not result in a stable improvement. On the contrary, it compromises the model's performance in most cases, particularly for Vesper-12. We hypothesize that the performance degradation may stem from the introduction of excessive supervision signals, which increases the difficulty of model training. Additionally, an abundance of supervision signals may constrain the model's flexibility, causing Vesper to closely resemble WavLM and consequently leading to a reduction in performance. Therefore, supervising the intermediate and final layers of Vesper is sufficient to achieve the optimal performance.

\subsection{Information Richness of the Last Layer Representation}

\begin{table}[t]
    \caption{Performance of WavLM and Vesper with different output representations fed to the classifier on the IEMOCAP, MELD, and CREMA-D datasets. $\Delta$ indicates an improvement (+) or a reduction (-). Last = representation of the last layer, Weighted = weighted summation of the representations of each layer, In. of C. = input of classifier}
    \label{tab_rep}
    \centering
    \begin{threeparttable}
    \begin{tabular}{c|c|c|ccc}
    \hline
    Dataset & Method & In. of C. &  WA & UA & WF1  \\ \hline
    \multirow{12}{*}{IEMOCAP} & \multirow{3}{*}{WavLM Base} & Weighted & 0.659\tnote{$\dagger$} & -  & - \\
    & & Last & 0.557 & 0.558 & 0.546 \\  \cline{3-6}
    & & $\Delta$ & -10.2\% & - & - \\ \cline{2-6}
    & \multirow{3}{*}{WavLM Large} & Weighted & 0.706\tnote{$\dagger$} & -  & - \\
    & & Last & 0.686 & 0.701 & 0.683 \\  \cline{3-6}
    & & $\Delta$ & -2.0\% & - & - \\ \cline{2-6}
    & \multirow{3}{*}{Vesper-4} & Weighted & 0.684 & 0.693 & 0.683 \\
    & & Last  & 0.681 & 0.699 & 0.679 \\  \cline{3-6}
    & & $\Delta$ & -0.3\% & +0.6\% & -0.4\% \\ \cline{2-6}
    & \multirow{3}{*}{Vesper-12} & Weighted & 0.707 & 0.708 & 0.706 \\
    & & Last & 0.705 & 0.712 & 0.705 \\  \cline{3-6}
    & & $\Delta$ & -0.2\% & +0.4\% & -0.1\% \\ \hline
    \multirow{12}{*}{MELD} & \multirow{3}{*}{WavLM Base} & Weighted & 0.499 & 0.201 & 0.400 \\
    & & Last & 0.481 & 0.143 & 0.313 \\ \cline{3-6}
    & & $\Delta$ & -1.8\% & -5.8\% & -8.7\% \\ \cline{2-6}
    & \multirow{3}{*}{WavLM Large} & Weighted & 0.542 & 0.253 & 0.476 \\
    & & Last  & 0.522 & 0.263 & 0.454 \\ \cline{3-6}
    & & $\Delta$ & -2.0\% & +1.0\% & -2.2\% \\ \cline{2-6}
    & \multirow{3}{*}{Vesper-4} & Weighted & 0.501 & 0.250 & 0.457 \\
    & & Last & 0.504 & 0.260 & 0.468 \\ \cline{3-6}
    & & $\Delta$ & +0.3\% & +1.0\% & +1.1\% \\ \cline{2-6}
    & \multirow{3}{*}{Vesper-12} & Weighted & 0.535 & 0.268 & 0.480 \\
    & & Last  & 0.526 & 0.266 & 0.476 \\ \cline{3-6}
    & & $\Delta$ & -0.9\% & -0.2\% & -0.4\% \\ \hline
    \multirow{12}{*}{CREMA-D} & \multirow{3}{*}{WavLM Base} & Weighted & 0.599 & 0.599 & 0.600 \\
    & & Last & 0.563 & 0.564 & 0.565 \\  \cline{3-6}
    & & $\Delta$ & -3.6\% & -3.5\% & -3.5\% \\ \cline{2-6}
    & \multirow{3}{*}{WavLM Large} & Weighted & 0.757 & 0.762 & 0.755 \\
    & & Last & 0.694 & 0.700 & 0.689 \\  \cline{3-6}
    & & $\Delta$ & -6.3\% & -6.2\% & -6.6\% \\ \cline{2-6}
    & \multirow{3}{*}{Vesper-4} & Weighted & 0.734 & `0.737 & 0.733 \\
    & & Last  & 0.719 & 0.723 & 0.714 \\  \cline{3-6}
    & & $\Delta$ & -1.5\% & -1.4\% & -1.9\% \\ \cline{2-6}
    & \multirow{3}{*}{Vesper-12} & Weighted & 0.772 & 0.776 & 0.768 \\
    & & Last & 0.774 & 0.778 & 0.772 \\  \cline{3-6}
    & & $\Delta$ & +0.2\% & +0.2\% & +0.4\% \\ \hline
    \end{tabular}
    \begin{tablenotes}
        \footnotesize
        \item[$\dagger$] Results from the official WavLM paper \cite{wavlm}.
    \end{tablenotes}
    \end{threeparttable}
\end{table}

Benefitting from the adoption of cross-layer self-supervision, the final output representation of Vesper contains both semantic information from the deep layers and acoustic information from the shallow layers. Hence, Vesper is expected to yield comparable performance when feeding only the representation of the last layer to the downstream classifier.
To validate this hypothesis,
we evaluate the performance when using last layer representation as input to the downstream classifier. 
The results are presented in Table~\ref{tab_rep}. On the IEMOCAP dataset, notable performance degradation is observed when only the representation from the last layer of WavLM is used. Specifically, there is a decrease of 10.2\% in WA for WavLM Base and 2.0\% in WA for WavLM Large. In contrast, Vesper using only the last layer representation displays only a minor decrease in performance (-0.2\%$\sim$-0.3\% in WA and -0.1\%$\sim$-0.4\% in WF1), and in the UA metric, it even exhibits improvement (+0.6\% for Vesper-4 and +0.4\% for Vesper-12). On the MELD dataset, using only the last layer representation from WavLM Base exhibits a severe decrease in performance (-1.8\% in WA, -5.8\% in UA, and -8.7\% in WF1). WavLM Large also exhibits a decrease of -2.2\% in the primary metric WF1. Remarkably, utilizing the last layer representation from Vesper-4 yields an improvement across all metrics, with a notable enhancement of +1.1\% in WF1. For Vesper-12 on the MELD dataset, the performance decrease caused by merely using the last layer representation is kept within the range of -0.2\% to -0.9\%. On the CREMA-D dataset, employing only the last layer representation of WavLM Base leads to a decrease of 3.5\%$\sim$3.6\% in all metrics. Similarly, using only the last layer representation of WavLM Large results in a decrease of 6.2\%$\sim$6.6\% in all evaluation metrics. For Vesper-4, solely using the last layer representation leads to a decrease of 1.4\%$\sim$1.9\% in all metrics, where the performance degradation is mitigated. When considering Vesper-12, the performance with the last layer representation increases by 0.2\%$\sim$0.4\%. Given the experimental results, we conclude that the last layer representation of Vesper is informative enough to perform speech emotion recognition. This characteristic simplifies the utilization of the pretrained model, as it is no longer necessary to extract representations from each layer individually. Instead, relying solely on the representation from the last layer of Vesper proves to be adequate and sufficient.

\subsection{Visualization of Deep Representations}

\begin{figure}[t]
\centering
\includegraphics[width=0.96\linewidth]{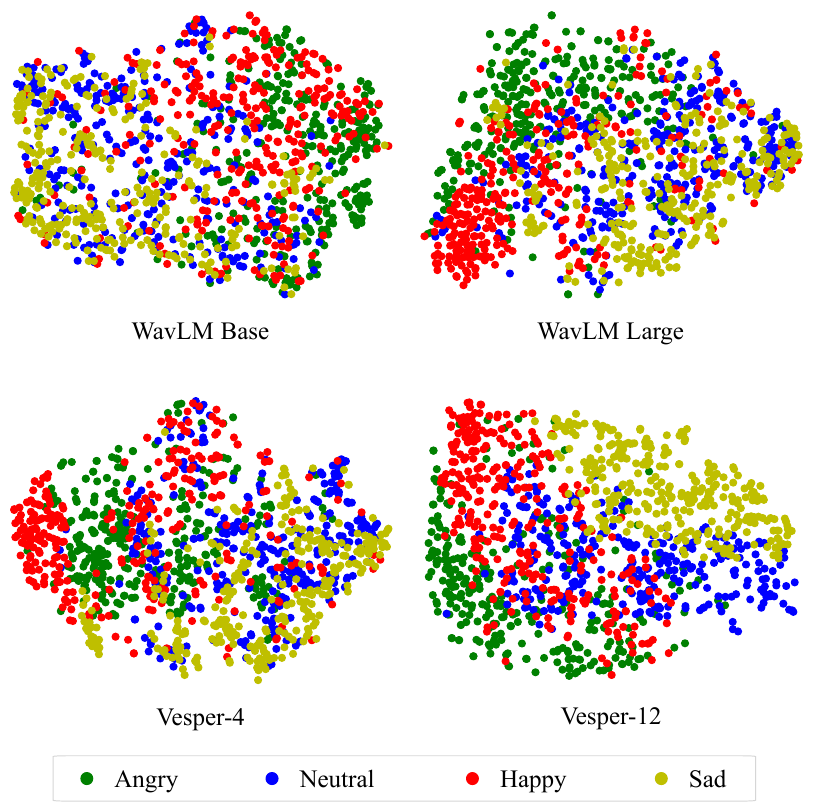}
\caption{Visualization of the deep representations from the last layer of WavLM Base, WavLM Large, Vesper-4, and Vesper-12 models on IEMOCAP.}
\label{visualize}
\end{figure}

To intuitively demonstrate Vesper's emotional specificity, we use the t-SNE \cite{t-sne} algorithm to visualize the deep representations from the last layer of the WavLM Base, WavLM Large, Vesper-4, and Vesper-12 models on the IEMOCAP dataset. The visualization results are illustrated in Fig.~\ref{visualize}. Similar results are observed on the CREMA-D dataset, which are not included in the paper to save space. Additionally, we omit the MELD dataset because the relatively poor performances of all models observed on the MELD dataset make the visualization indistinguishable. The visualization results show that the representations of all samples obtained by WavLM Base are mixed together and completely indistinguishable from each other. Our Vesper-4 demonstrates the capability to differentiate between happy and sad samples and differentiate between angry and sad samples to some extent. However, the representations of happy and angry emotions remain confusing. Similarities in acoustic features between happy and angry samples, such as high energy and a fast speech rate, present a challenge for recognition. Additionally, the neutral and happy samples exhibit better clustering compared to WavLM Base. In WavLM Large, the angry, happy, and sad samples can be approximately distinguished. However, the neutral representations from WavLM Large are still severely confused with other emotional representations. This issue is effectively solved to some extent by Vesper-12. Although there is some overlap in the margins of different emotional samples, Vesper-12 successfully distinguishes most samples in terms of the four different emotional states. The visualization results once again confirm the effectiveness of the proposed Vesper.

\begin{table}[t]
    \caption{Performance of HuBERT and Vesper on the IEMOCAP, MELD, and CREMA-D datasets for speech emotion recognition}
    \label{tab_hubert}
    \centering
    \begin{threeparttable}
    \begin{tabular}{c|c|ccc}
    \hline
    Dataset & Method & WA & UA & WF1  \\ \hline
    \multirow{4}{*}{IEMOCAP} & HuBERT Base  & 0.649\tnote{$\dagger$} & -  & -  \\
    & \tnote{$\ddagger$}\; Vesper-4    & \textbf{0.667} & \textbf{0.670} & \textbf{0.653}   \\ \cline{2-5}
    & HuBERT Large & 0.676\tnote{$\dagger$} & - & - \\
    & \tnote{$\ddagger$}\; Vesper-12   & \textbf{0.688} & \textbf{0.699} & \textbf{0.686} \\ \hline
    \multirow{4}{*}{MELD} & HuBERT Base & 0.483 & 0.193 & 0.386   \\
    & \tnote{$\ddagger$}\; Vesper-4    & \textbf{0.498} & \textbf{0.241} & \textbf{0.443}   \\ \cline{2-5}
    & HuBERT Large & \textbf{0.515} & 0.206 & 0.422  \\
    & \tnote{$\ddagger$}\; Vesper-12   & 0.509 & \textbf{0.248} & \textbf{0.448}  \\ \hline
    \multirow{4}{*}{CREMA-D} & HuBERT Base & 0.614 & 0.617 & 0.610   \\
    & \tnote{$\ddagger$}\; Vesper-4    & \textbf{0.721} & \textbf{0.723} & \textbf{0.720}   \\ \cline{2-5}
    & HuBERT Large & 0.745 & 0.743 & 0.750 \\
    & \tnote{$\ddagger$}\; Vesper-12   & \textbf{0.762} & \textbf{0.765} & \textbf{0.761}  \\ \hline
    \end{tabular}
    \begin{tablenotes}
        \footnotesize
        \item[$\dagger$] Results from the SUPERB benchmark \cite{superb}.
        \item[$\ddagger$] Vesper is initialized and supervised by HuBERT Large \cite{hubert}.
    \end{tablenotes}
    \end{threeparttable}
\end{table}

\subsection{Replacing WavLM with HuBERT}

To further validate the feasibility of the proposed paradigm, we replace WavLM with HuBERT \cite{hubert}, another widely utilized large-scale pretrained model in the speech domain. Similar to WavLM, HuBERT is available in two versions: Base and Large, comprising 12 and 24 layers respectively. In this subsection, Vesper is initialized with the parameters of HuBERT Large and guided by the intermediate and final outputs of HuBERT Large through label-free self-supervision. Subsequently, we compare the performances of the resulting Vesper with the original HuBERT model on the used datasets. As shown in Table~\ref{tab_hubert}, our Vesper-4 with 4 layers consistently outperforms HuBERT Base with 12 layers across all three datasets. Specifically, our Vesper-4 achieves a 1.8\% gain in WA over HuBERT Base on the IEMOCAP dataset, a 5.7\% gain in WF1 on the MELD dataset, and a 10.7\% gain in WA on the CREMA-D dataset. Similar results are observed when comparing the 12-layer Vesper-12 with the 24-layer HuBERT Large. Across all datasets, Vesper-12 demonstrates significantly superior performance to HuBERT Large in all metrics, with the exception of the less crucial WA metric in the MELD dataset, where HuBERT Large slightly outperforms Vesper-12. The results suggest that the proposed paradigm is also effective when applied to the pretrained HuBERT model, confirming the feasibility and universality of our paradigm.

\section{Conclusion}

In this paper, we propose a new paradigm to generate task-specific pretrained models for speech emotion recognition by applying compression and label-free adaptation. We produce an improved emotion-specific pretrained encoder called Vesper. Building upon the large-scale pretrained WavLM, Vesper directly utilizes the parameters of WavLM for initialization and improves the specificity for emotion recognition by employing an emotion-guided masking strategy during emotion-specific pretraining. In addition, hierarchical self-supervision and cross-layer self-supervision are proposed to learn the acoustic and semantic information embedded in speech signals. Experimental results on three emotion recognition corpora demonstrate that our Vesper with 4 layers substantially outperforms WavLM Base with 12 layers, and the performance of Vesper with 12 layers exceeds that of WavLM Large with 24 layers. Additionally, ablation experiments are conducted to analyze the individual contributions of each component. In the future, we intend to develop task-specific pretrained models for other tasks, including speaker recognition and speech enhancement, by using the paradigm presented in this paper and redesigning the task-specific objectives and training strategies based on the characteristics of the target tasks.

\section*{Acknowledgments}
The work was supported by the National Key R\&D Program of China 2022YFB4500600;
in part by the Science and Technology Project of Guangzhou 202103010002;
in part by Natural Science Foundation of Guangdong Province 2022A1515011588;
in part by Nansha Key Project 2022ZD011;
in part by the Fundamental Research Funds for the Central Universities 2022ZYGXZR0075;
in part by Key-Area Research and Development Program of Guangdong Province 2022B0101010003;
in part by the Guangdong Provincial Key Laboratory of Human Digital Twin 2022B1212010004;
and in part by Key Laboratory of Cognitive Radio and Information Processing (Ministry of Education, Guilin University of Electronic Technology).

\bibliographystyle{IEEEtran}
\bibliography{refs}

\begin{thebibliography}{10}
\providecommand{\url}[1]{#1}
\csname url@samestyle\endcsname
\providecommand{\newblock}{\relax}
\providecommand{\bibinfo}[2]{#2}
\providecommand{\BIBentrySTDinterwordspacing}{\spaceskip=0pt\relax}
\providecommand{\BIBentryALTinterwordstretchfactor}{4}
\providecommand{\BIBentryALTinterwordspacing}{\spaceskip=\fontdimen2\font plus
\BIBentryALTinterwordstretchfactor\fontdimen3\font minus
  \fontdimen4\font\relax}
\providecommand{\BIBforeignlanguage}[2]{{%
\expandafter\ifx\csname l@#1\endcsname\relax
\typeout{** WARNING: IEEEtran.bst: No hyphenation pattern has been}%
\typeout{** loaded for the language `#1'. Using the pattern for}%
\typeout{** the default language instead.}%
\else
\language=\csname l@#1\endcsname
\fi
#2}}
\providecommand{\BIBdecl}{\relax}
\BIBdecl

\bibitem{survey_speech}
S.~Liu, A.~Mallol-Ragolta, E.~Parada-Cabaleiro, K.~Qian, X.~Jing, A.~Kathan,
  B.~Hu, and B.~W. Schuller, ``Audio self-supervised learning: A survey,''
  \emph{Patterns}, vol.~3, no.~12, p. 100616, 2022.

\bibitem{survey}
Y.~Du, Z.~Liu, J.~Li, and W.~X. Zhao, ``A survey of vision-language pre-trained
  models,'' \emph{arXiv preprint arXiv:2202.10936}, 2022.

\bibitem{gpt-3}
T.~Brown, B.~Mann, N.~Ryder, M.~Subbiah, J.~D. Kaplan, P.~Dhariwal,
  A.~Neelakantan, P.~Shyam, G.~Sastry, A.~Askell, S.~Agarwal, A.~Herbert-Voss,
  G.~Krueger, T.~Henighan, R.~Child, A.~Ramesh, D.~Ziegler, J.~Wu, C.~Winter,
  C.~Hesse, M.~Chen, E.~Sigler, M.~Litwin, S.~Gray, B.~Chess, J.~Clark,
  C.~Berner, S.~McCandlish, A.~Radford, I.~Sutskever, and D.~Amodei, ``Language
  models are few-shot learners,'' in \emph{Advances in Neural Information
  Processing Systems}, vol.~33, 2020, pp. 1877--1901.

\bibitem{bert}
J.~Devlin, M.-W. Chang, K.~Lee, and K.~Toutanova, ``{BERT}: Pre-training of
  deep bidirectional transformers for language understanding,'' \emph{arXiv
  preprint arXiv:1810.04805}, 2018.

\bibitem{clip}
A.~Radford, J.~W. Kim, C.~Hallacy, A.~Ramesh, G.~Goh, S.~Agarwal, G.~Sastry,
  A.~Askell, P.~Mishkin, J.~Clark \emph{et~al.}, ``Learning transferable visual
  models from natural language supervision,'' in \emph{International conference
  on machine learning}, 2021, pp. 8748--8763.

\bibitem{wav2vec}
S.~Schneider, A.~Baevski, R.~Collobert, and M.~Auli, ``wav2vec: Unsupervised
  pre-training for speech recognition,'' in \emph{Interspeech}, 2019, pp.
  3465--3469.

\bibitem{wav2vec2}
A.~Baevski, Y.~Zhou, A.~Mohamed, and M.~Auli, ``wav2vec 2.0: A framework for
  self-supervised learning of speech representations,'' in \emph{Advances in
  Neural Information Processing Systems}, vol.~33, 2020, pp. 12\,449--12\,460.

\bibitem{hubert}
W.-N. Hsu, B.~Bolte, Y.-H.~H. Tsai, K.~Lakhotia, R.~Salakhutdinov, and
  A.~Mohamed, ``{HuBERT}: Self-supervised speech representation learning by
  masked prediction of hidden units,'' \emph{IEEE/ACM Transactions on Audio,
  Speech, and Language Processing}, vol.~29, pp. 3451--3460, 2021.

\bibitem{wavlm}
S.~Chen, C.~Wang, Z.~Chen, Y.~Wu, S.~Liu, Z.~Chen, J.~Li, N.~Kanda,
  T.~Yoshioka, X.~Xiao, J.~Wu, L.~Zhou, S.~Ren, Y.~Qian, Y.~Qian, J.~Wu,
  M.~Zeng, X.~Yu, and F.~Wei, ``{WavLM}: Large-scale self-supervised
  pre-training for full stack speech processing,'' \emph{IEEE Journal of
  Selected Topics in Signal Processing}, vol.~16, no.~6, pp. 1505--1518, 2022.

\bibitem{llama}
H.~Touvron, T.~Lavril, G.~Izacard, X.~Martinet, M.-A. Lachaux, T.~Lacroix,
  B.~Rozi{\`e}re, N.~Goyal, E.~Hambro, F.~Azhar, A.~Rodriguez, A.~Joulin,
  E.~Grave, and G.~Lample, ``Llama: Open and efficient foundation language
  models,'' \emph{arXiv preprint arXiv:2302.13971}, 2023.

\bibitem{Palm}
A.~Chowdhery, S.~Narang, J.~Devlin, M.~Bosma, G.~Mishra, A.~Roberts, P.~Barham,
  H.~W. Chung, C.~Sutton, S.~Gehrmann, P.~Schuh, K.~Shi, S.~Tsvyashchenko,
  J.~Maynez, A.~Rao, P.~Barnes, Y.~Tay, N.~Shazeer, V.~Prabhakaran, E.~Reif,
  N.~Du, B.~Hutchinson, R.~Pope, J.~Bradbury, J.~Austin, M.~Isard, G.~Gur-Ari,
  P.~Yin, T.~Duke, A.~Levskaya, S.~Ghemawat, S.~Dev, H.~Michalewski, X.~Garcia,
  V.~Misra, K.~Robinson, L.~Fedus, D.~Zhou, D.~Ippolito, D.~Luan, H.~Lim,
  B.~Zoph, A.~Spiridonov, R.~Sepassi, D.~Dohan, S.~Agrawal, M.~Omernick, A.~M.
  Dai, T.~S. Pillai, M.~Pellat, A.~Lewkowycz, E.~Moreira, R.~Child, O.~Polozov,
  K.~Lee, Z.~Zhou, X.~Wang, B.~Saeta, M.~Diaz, O.~Firat, M.~Catasta, J.~Wei,
  K.~Meier-Hellstern, D.~Eck, J.~Dean, S.~Petrov, and N.~Fiedel, ``Palm:
  Scaling language modeling with pathways,'' \emph{arXiv preprint
  arXiv:2204.02311}, 2022.

\bibitem{ser_with_pre}
L.-W. Chen and A.~Rudnicky, ``Exploring wav2vec 2.0 fine tuning for improved
  speech emotion recognition,'' in \emph{IEEE International Conference on
  Acoustics, Speech and Signal Processing}, 2023, pp. 1--5.

\bibitem{qa_with_pre}
Q.~Cao, H.~Trivedi, A.~Balasubramanian, and N.~Balasubramanian, ``{DeFormer}:
  Decomposing pre-trained transformers for faster question answering,'' in
  \emph{Proceedings of the 58th Annual Meeting of the Association for
  Computational Linguistics}, 2020, pp. 4487--4497.

\bibitem{adapter}
R.~Weng, H.~Yu, S.~Huang, S.~Cheng, and W.~Luo, ``Acquiring knowledge from
  pre-trained model to neural machine translation,'' in \emph{Proceedings of
  the AAAI conference on artificial intelligence}, vol.~34, no.~05, 2020, pp.
  9266--9273.

\bibitem{as_with_pre}
A.~R. Fabbri, S.~Han, H.~Li, H.~Li, M.~Ghazvininejad, S.~Joty, D.~Radev, and
  Y.~Mehdad, ``Improving zero and few-shot abstractive summarization with
  intermediate fine-tuning and data augmentation,'' in \emph{Proceedings of the
  2021 Conference of the North American Chapter of the Association for
  Computational Linguistics: Human Language Technologies}, 2021, pp. 704--717.

\bibitem{distilling}
R.~Tang, Y.~Lu, L.~Liu, L.~Mou, O.~Vechtomova, and J.~Lin, ``Distilling
  task-specific knowledge from {BERT} into simple neural networks,''
  \emph{arXiv preprint arXiv:1903.12136}, 2019.

\bibitem{adversarial}
M.~Zhang, N.~U. Naresh, and Y.~He, ``Adversarial data augmentation for
  task-specific knowledge distillation of pre-trained transformers,'' in
  \emph{Proceedings of the AAAI Conference on Artificial Intelligence},
  vol.~36, no.~10, 2022, pp. 11\,685--11\,693.

\bibitem{DistilBERT}
V.~Sanh, L.~Debut, J.~Chaumond, and T.~Wolf, ``{DistilBERT}, a distilled
  version of bert: smaller, faster, cheaper and lighter,'' \emph{arXiv preprint
  arXiv:1910.01108}, 2019.

\bibitem{AdaBERT}
D.~Chen, Y.~Li, M.~Qiu, Z.~Wang, B.~Li, B.~Ding, H.~Deng, J.~Huang, W.~Lin, and
  J.~Zhou, ``Adabert: Task-adaptive {BERT} compression with differentiable
  neural architecture search,'' in \emph{Proceedings of the Twenty-Ninth
  International Joint Conference on Artificial Intelligence}, 2021, pp.
  2463--2469.

\bibitem{12layer}
J.~Shah, Y.~K. Singla, C.~Chen, and R.~R. Shah, ``What all do audio transformer
  models hear? {Probing} acoustic representations for language delivery and its
  structure,'' \emph{arXiv preprint arXiv:2101.00387}, 2021.

\bibitem{A_and_S}
P.~Tzirakis, A.~Nguyen, S.~Zafeiriou, and B.~W. Schuller, ``Speech emotion
  recognition using semantic information,'' in \emph{IEEE International
  Conference on Acoustics, Speech and Signal Processing}, 2021, pp. 6279--6283.

\bibitem{A_and_S_2}
B.~T. Atmaja, A.~Sasou, and M.~Akagi, ``Survey on bimodal speech emotion
  recognition from acoustic and linguistic information fusion,'' \emph{Speech
  Communication}, vol. 140, pp. 11--28, 2022.

\bibitem{roberta}
Y.~Liu, M.~Ott, N.~Goyal, J.~Du, M.~Joshi, D.~Chen, O.~Levy, M.~Lewis,
  L.~Zettlemoyer, and V.~Stoyanov, ``{RoBERTa}: A robustly optimized {BERT}
  pretraining approach,'' \emph{arXiv preprint arXiv:1907.11692}, 2019.

\bibitem{t5}
C.~Raffel, N.~Shazeer, A.~Roberts, K.~Lee, S.~Narang, M.~Matena, Y.~Zhou,
  W.~Li, and P.~J. Liu, ``Exploring the limits of transfer learning with a
  unified text-to-text transformer,'' \emph{Journal of Machine Learning
  Research}, vol.~21, no.~1, pp. 5485--5551, 2020.

\bibitem{ner_with_pre}
U.~Naseem, M.~Khushi, V.~Reddy, S.~Rajendran, I.~Razzak, and J.~Kim,
  ``Bioalbert: A simple and effective pre-trained language model for biomedical
  named entity recognition,'' in \emph{International Joint Conference on Neural
  Networks}, 2021, pp. 1--7.

\bibitem{mt_with_pre}
G.~Chen, S.~Ma, Y.~Chen, L.~Dong, D.~Zhang, J.~Pan, W.~Wang, and F.~Wei,
  ``Zero-shot cross-lingual transfer of neural machine translation with
  multilingual pretrained encoders,'' in \emph{Proceedings of the 2021
  Conference on Empirical Methods in Natural Language Processing}, 2021, pp.
  15--26.

\bibitem{moco}
K.~He, H.~Fan, Y.~Wu, S.~Xie, and R.~Girshick, ``Momentum contrast for
  unsupervised visual representation learning,'' in \emph{Proceedings of the
  IEEE/CVF Conference on Computer Vision and Pattern Recognition}, June 2020.

\bibitem{vit}
A.~Dosovitskiy, L.~Beyer, A.~Kolesnikov, D.~Weissenborn, X.~Zhai,
  T.~Unterthiner, M.~Dehghani, M.~Minderer, G.~Heigold, S.~Gelly, J.~Uszkoreit,
  and N.~Houlsby, ``An image is worth 16x16 words: Transformers for image
  recognition at scale,'' in \emph{International Conference on Learning
  Representations}, 2021, pp. 1--21.

\bibitem{simclr}
T.~Chen, S.~Kornblith, M.~Norouzi, and G.~Hinton, ``A simple framework for
  contrastive learning of visual representations,'' in \emph{International
  Conference on Machine Learning}, 2020, pp. 1597--1607.

\bibitem{visual_1}
X.~Sun, P.~Chen, L.~Chen, C.~Li, T.~H. Li, M.~Tan, and C.~Gan, ``Masked motion
  encoding for self-supervised video representation learning,'' in
  \emph{Proceedings of the IEEE/CVF Conference on Computer Vision and Pattern
  Recognition}, 2023, pp. 2235--2245.

\bibitem{visual_2}
P.~Chen, D.~Huang, D.~He, X.~Long, R.~Zeng, S.~Wen, M.~Tan, and C.~Gan,
  ``Rspnet: Relative speed perception for unsupervised video representation
  learning,'' in \emph{Proceedings of the AAAI Conference on Artificial
  Intelligence}, vol.~35, no.~2, 2021, pp. 1045--1053.

\bibitem{od_with_pre}
J.~Beal, E.~Kim, E.~Tzeng, D.~H. Park, A.~Zhai, and D.~Kislyuk, ``Toward
  transformer-based object detection,'' \emph{arXiv preprint arXiv:2012.09958},
  2020.

\bibitem{is_with_pre}
S.~Minaee, Y.~Boykov, F.~Porikli, A.~Plaza, N.~Kehtarnavaz, and D.~Terzopoulos,
  ``Image segmentation using deep learning: A survey,'' \emph{IEEE Transactions
  on Pattern Analysis and Machine Intelligence}, vol.~44, no.~7, pp.
  3523--3542, 2022.

\bibitem{ic_with_pre}
J.~Chen, H.~Guo, K.~Yi, B.~Li, and M.~Elhoseiny, ``{VisualGPT}: Data-efficient
  adaptation of pretrained language models for image captioning,'' in
  \emph{Proceedings of the IEEE/CVF Conference on Computer Vision and Pattern
  Recognition}, June 2022, pp. 18\,030--18\,040.

\bibitem{sam}
A.~Kirillov, E.~Mintun, N.~Ravi, H.~Mao, C.~Rolland, L.~Gustafson, T.~Xiao,
  S.~Whitehead, A.~C. Berg, W.-Y. Lo \emph{et~al.}, ``Segment anything,''
  \emph{arXiv preprint arXiv:2304.02643}, 2023.

\bibitem{flamingo}
J.-B. Alayrac, J.~Donahue, P.~Luc, A.~Miech, I.~Barr, Y.~Hasson, K.~Lenc,
  A.~Mensch, K.~Millican, M.~Reynolds \emph{et~al.}, ``Flamingo: a visual
  language model for few-shot learning,'' \emph{Advances in Neural Information
  Processing Systems}, vol.~35, pp. 23\,716--23\,736, 2022.

\bibitem{dall-e}
A.~Ramesh, M.~Pavlov, G.~Goh, S.~Gray, C.~Voss, A.~Radford, M.~Chen, and
  I.~Sutskever, ``Zero-shot text-to-image generation,'' in \emph{International
  Conference on Machine Learning}, 2021, pp. 8821--8831.

\bibitem{asr_pre}
Q.-S. Zhu, J.~Zhang, Z.-Q. Zhang, and L.-R. Dai, ``A joint speech enhancement
  and self-supervised representation learning framework for noise-robust speech
  recognition,'' \emph{IEEE/ACM Transactions on Audio, Speech, and Language
  Processing}, vol.~31, pp. 1927--1939, 2023.

\bibitem{se_with_pre}
X.-Y. Zhao, Q.-S. Zhu, and J.~Zhang, ``Speech enhancement using self-supervised
  pre-trained model and vector quantization,'' in \emph{Asia-Pacific Signal and
  Information Processing Association Annual Summit and Conference}, 2022, pp.
  330--334.

\bibitem{se_with_pre_2}
R.~E. Zezario, S.-W. Fu, F.~Chen, C.-S. Fuh, H.-M. Wang, and Y.~Tsao, ``Deep
  learning-based non-intrusive multi-objective speech assessment model with
  cross-domain features,'' \emph{IEEE/ACM Transactions on Audio, Speech, and
  Language Processing}, vol.~31, pp. 54--70, 2023.

\bibitem{dst}
W.~Chen, X.~Xing, X.~Xu, J.~Pang, and L.~Du, ``{DST}: Deformable speech
  transformer for emotion recognition,'' in \emph{IEEE International Conference
  on Acoustics, Speech and Signal Processing}, 2023, pp. 1--5.

\bibitem{shiftser}
S.~Shen, F.~Liu, and A.~Zhou, ``{Mingling or misalignment? Temporal shift for
  speech emotion recognition with pre-trained representations},'' in \emph{IEEE
  International Conference on Acoustics, Speech and Signal Processing}, 2023,
  pp. 1--5.

\bibitem{tac_1}
J.-L. Li and C.-C. Lee, ``An enroll-to-verify approach for cross-task unseen
  emotion class recognition,'' \emph{IEEE Transactions on Affective Computing},
  to be published, doi: 10.1109/TAFFC.2022.3183166.

\bibitem{finetune}
Y.~Su, X.~Han, Y.~Lin, Z.~Zhang, Z.~Liu, P.~Li, J.~Zhou, and M.~Sun, ``Css-lm:
  A contrastive framework for semi-supervised fine-tuning of pre-trained
  language models,'' \emph{IEEE/ACM Transactions on Audio, Speech, and Language
  Processing}, vol.~29, pp. 2930--2941, 2021.

\bibitem{fastbert}
W.~Liu, P.~Zhou, Z.~Wang, Z.~Zhao, H.~Deng, and Q.~Ju, ``{F}ast{BERT}: a
  self-distilling {BERT} with adaptive inference time,'' in \emph{Proceedings
  of the 58th Annual Meeting of the Association for Computational Linguistics},
  2020, pp. 6035--6044.

\bibitem{DAAE}
Y.~Gao, J.~Liu, L.~Wang, and J.~Dang, ``Domain-adversarial autoencoder with
  attention based feature level fusion for speech emotion recognition,'' in
  \emph{IEEE International Conference on Acoustics, Speech and Signal
  Processing}, 2021, pp. 6314--6318.

\bibitem{LanSER}
T.~Gong, J.~Belanich, K.~Somandepalli, A.~Nagrani, B.~Eoff, and B.~Jou,
  ``{LanSER: L}anguage-model supported speech emotion recognition,'' in
  \emph{INTERSPEECH}, 2023, pp. 2408--2412.

\bibitem{ksT}
W.~Chen, X.~Xing, X.~Xu, J.~Yang, and J.~Pang, ``Key-sparse transformer for
  multimodal speech emotion recognition,'' in \emph{IEEE International
  Conference on Acoustics, Speech and Signal Processing}, 2022, pp. 6897--6901.

\bibitem{rahman}
M.~Rahman, Y.~Cao, X.~Sun, B.~Li, and Y.~Hao, ``Deep pre-trained networks as a
  feature extractor with xgboost to detect tuberculosis from chest x-ray,''
  \emph{Computers \& Electrical Engineering}, vol.~93, p. 107252, 2021.

\bibitem{extractor}
C.-C. Kuo, K.-Y. Chen, and S.-B. Luo, ``Audio-aware spoken multiple-choice
  question answering with pre-trained language models,'' \emph{IEEE/ACM
  Transactions on Audio, Speech, and Language Processing}, vol.~29, pp.
  3170--3179, 2021.

\bibitem{CA-MSER}
H.~Zou, Y.~Si, C.~Chen, D.~Rajan, and E.~S. Chng, ``Speech emotion recognition
  with co-attention based multi-level acoustic information,'' in \emph{IEEE
  International Conference on Acoustics, Speech and Signal Processing}, 2022,
  pp. 7367--7371.

\bibitem{Phukan}
O.~{Chetia Phukan}, A.~{Balaji Buduru}, and R.~Sharma, ``Transforming the
  embeddings: {A} lightweight technique for speech emotion recognition tasks,''
  in \emph{INTERSPEECH}, 2023, pp. 1903--1907.

\bibitem{wu}
T.-Y. Wu, C.-A. Li, T.-H. Lin, T.-Y. Hsu, and H.-Y. Lee, ``The ability of
  self-supervised speech models for audio representations,'' \emph{arXiv
  preprint arXiv:2209.12900}, 2022.

\bibitem{Transformer}
A.~Vaswani, N.~Shazeer, N.~Parmar, J.~Uszkoreit, L.~Jones, A.~N. Gomez,
  L.~Kaiser, and I.~Polosukhin, ``Attention is all you need,'' in
  \emph{Proceedings of the 31st International Conference on Neural Information
  Processing Systems}, 2017, pp. 5998--6008.

\bibitem{relu}
X.~Glorot, A.~Bordes, and Y.~Bengio, ``Deep sparse rectifier neural networks,''
  in \emph{Proceedings of the Fourteenth International Conference on Artificial
  Intelligence and Statistics}, 2011, pp. 315--323.

\bibitem{superb}
S.~wen Yang, P.-H. Chi, Y.-S. Chuang, C.-I.~J. Lai, K.~Lakhotia, Y.~Y. Lin,
  A.~T. Liu, J.~Shi, X.~Chang, G.-T. Lin, T.-H. Huang, W.-C. Tseng, K.~tik Lee,
  D.-R. Liu, Z.~Huang, S.~Dong, S.-W. Li, S.~Watanabe, A.~Mohamed, and
  H.~yi~Lee, ``{SUPERB: S}peech processing universal performance benchmark,''
  in \emph{Interspeech}, 2021, pp. 1194--1198.

\bibitem{energy_matter}
O.-W. Kwon, K.~Chan, J.~Hao, and T.-W. Lee, ``Emotion recognition by speech
  signals,'' in \emph{Interspeech}, 2003, pp. 125--128.

\bibitem{energy_matter_2}
K.~Venkataramanan and H.~R. Rajamohan, ``Emotion recognition from speech,''
  \emph{arXiv preprint arXiv:1912.10458}, 2019.

\bibitem{energy_matter_3}
M.~El~Ayadi, M.~S. Kamel, and F.~Karray, ``Survey on speech emotion
  recognition: Features, classification schemes, and databases,'' \emph{Pattern
  recognition}, vol.~44, no.~3, pp. 572--587, 2011.

\bibitem{speechformer++}
W.~Chen, X.~Xing, X.~Xu, J.~Pang, and L.~Du, ``{SpeechFormer++}: A hierarchical
  efficient framework for paralinguistic speech processing,'' \emph{IEEE/ACM
  Transactions on Audio, Speech, and Language Processing}, vol.~31, pp.
  775--788, 2023.

\bibitem{lssed}
W.~Fan, X.~Xu, X.~Xing, W.~Chen, and D.~Huang, ``{LSSED}: a large-scale dataset
  and benchmark for speech emotion recognition,'' in \emph{IEEE International
  Conference on Acoustics, Speech and Signal Processing}, 2021, pp. 641--645.

\bibitem{IEMOCAP}
C.~Busso, M.~Bulut, C.-C. Lee, A.~Kazemzadeh, E.~Mower, S.~Kim, J.~N. Chang,
  S.~Lee, and S.~S. Narayanan, ``{IEMOCAP: Interactive emotional dyadic motion
  capture database},'' \emph{Language Resources and Evaluation}, vol.~42,
  no.~4, pp. 335--359, 2008.

\bibitem{meld}
S.~Poria, D.~Hazarika, N.~Majumder, G.~Naik, E.~Cambria, and R.~Mihalcea,
  ``{MELD: A} multimodal multi-party dataset for emotion recognition in
  conversations,'' \emph{arXiv preprint arXiv:1810.02508}, 2019.

\bibitem{crema}
H.~Cao, D.~G. Cooper, M.~K. Keutmann, R.~C. Gur, A.~Nenkova, and R.~Verma,
  ``{CREMA-D}: Crowd-sourced emotional multimodal actors dataset,'' \emph{IEEE
  Transactions on Affective Computing}, vol.~5, no.~4, pp. 377--390, 2014.

\bibitem{adamw}
I.~Loshchilov and F.~Hutter, ``Decoupled weight decay regularization,''
  \emph{arXiv preprint arXiv:1711.05101}, 2017.

\bibitem{sgd}
H.~Robbins and S.~Monro, ``A stochastic approximation method,'' \emph{The
  annals of mathematical statistics}, pp. 400--407, 1951.

\bibitem{cos_decay}
I.~Loshchilov and F.~Hutter, ``Sgdr: Stochastic gradient descent with warm
  restarts,'' \emph{arXiv preprint arXiv:1608.03983}, 2016.

\bibitem{speechformer}
W.~Chen, X.~Xing, X.~Xu, J.~Pang, and L.~Du, ``{SpeechFormer}: A hierarchical
  efficient framework incorporating the characteristics of speech,'' in
  \emph{Interspeech}, 2022, pp. 346--350.

\bibitem{MCFN}
X.~Zhang and Y.~Li, ``A dual attention-based modality-collaborative fusion
  network for emotion recognition,'' in \emph{INTERSPEECH}, 2023, pp.
  1468--1472.

\bibitem{ISNet}
W.~Fan, X.~Xu, B.~Cai, and X.~Xing, ``{ISNet}: Individual standardization
  network for speech emotion recognition,'' \emph{IEEE/ACM Transactions on
  Audio, Speech, and Language Processing}, vol.~30, pp. 1803--1814, 2022.

\bibitem{DECN}
Z.~Lian, B.~Liu, and J.~Tao, ``{DECN}: Dialogical emotion correction network
  for conversational emotion recognition,'' \emph{Neurocomputing}, vol. 454,
  pp. 483--495, 2021.

\bibitem{SCFA}
H.~Zhao, B.~Li, and Z.~Zhang, ``Speaker-aware cross-modal fusion architecture
  for conversational emotion recognition,'' in \emph{INTERSPEECH}, 2023, pp.
  2718--2722.

\bibitem{IG-CNN}
E.~Jing, Y.~Liu, Y.~Chai, J.~Sun, S.~Samtani, Y.~Jiang, and Y.~Qian, ``A deep
  interpretable representation learning method for speech emotion
  recognition,'' \emph{Information Processing \& Management}, vol.~60, no.~6,
  p. 103501, 2023.

\bibitem{t-sne}
L.~Van~der Maaten and G.~Hinton, ``Visualizing data using t-sne.''
  \emph{Journal of Machine Learning Research}, vol.~9, no.~11, pp. 2579--2605,
  2008.

\end{thebibliography}


\vspace{11pt}
\vspace{-33pt}
\begin{IEEEbiography}[{\includegraphics[width=1in,height=1.25in,clip,keepaspectratio]{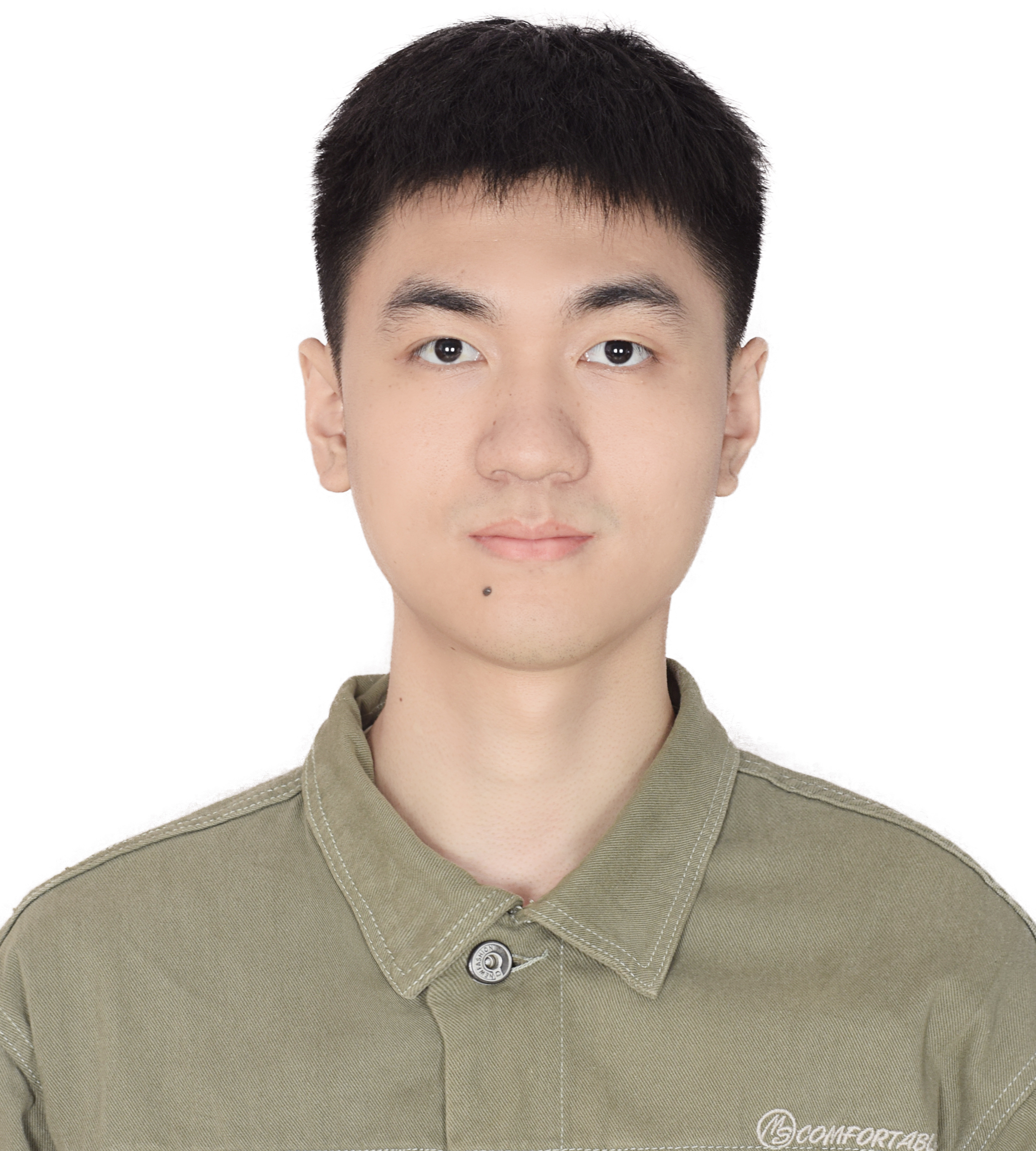}}]{Weidong Chen} (Student Member, IEEE) received the B.E. degree in electronic science and technology from South China University of Technology, Guangzhou, China, in 2021. He is currently working toward the M.E. degree with the School of Electronic and Information Engineering, South China University of Technology. His research interests include speech emotion recognition, multimodal emotion recognition, and deep learning in speech processing.
\end{IEEEbiography}

\vspace{11pt}
\vspace{-33pt}
\begin{IEEEbiography}[{\includegraphics[width=1in,height=1.25in,clip,keepaspectratio]{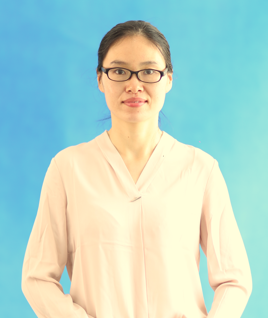}}]{Xiaofen Xing} (Member, IEEE) received the B.S., M.S., and Ph.D. degrees from South China University of Technology, Guangzhou, China, in 2001, 2004 and 2013, respectively. Since 2017, she has been an Associate Professor with the School of Electronic and Information Engineering, South China University of Technology. Her main research interests include speech emotion analysis, image/video processing, and human computer interaction.
\end{IEEEbiography}

\vspace{11pt}
\vspace{-33pt}
\begin{IEEEbiography}[{\includegraphics[width=1in,height=1.25in,clip,keepaspectratio]{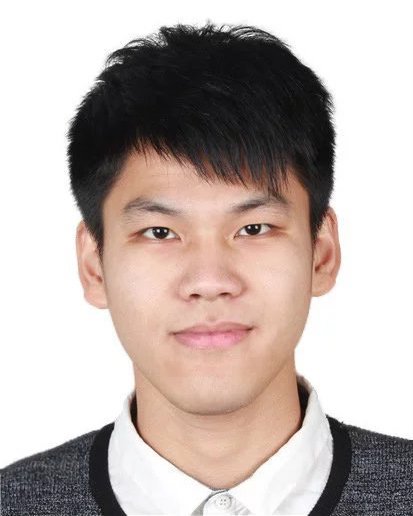}}]{Peihao Chen} received the B.E. degree in Automation Science and Engineering from South China University of Technology, China, in 2018. He is working toward the Ph.D. degree in the School of Software Engineering, South China University of Technology, China. His research interests include Deep Learning in Video and Audio Understanding.
\end{IEEEbiography}

\vspace{11pt}
\vspace{-33pt}
\begin{IEEEbiography}[{\includegraphics[width=1in,height=1.25in,clip,keepaspectratio]{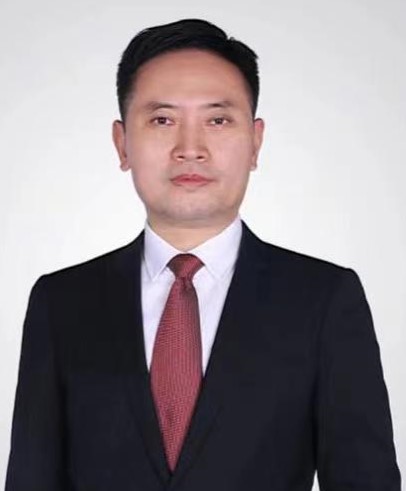}}]{Xiangmin Xu} (Senior Member, IEEE) received the Ph.D. degree from South China University of Technology, Guangzhou, China. He is currently a Full Professor with the School of Electronic and Information Engineering, South China University of Technology. His research interests include image/video processing, human computer interaction, computer vision, and machine learning.
\end{IEEEbiography}

\vfill

\end{document}